\documentclass[showpacs,prb,aps,supersc-riptaddress,twocolumn,floatfix]{revtex4-1}
\usepackage[T1]{fontenc}
\usepackage{graphicx}
\usepackage{amsmath,float}
\usepackage{amsfonts}
\usepackage{amsthm}
\usepackage{amssymb}
\usepackage{multirow}
\usepackage[a4paper,left=1cm,right=1cm,top=2cm,bottom=2cm]{geometry}
\usepackage{bbm}
\usepackage{color}
\newcommand{\K}{{\bf K}}
\newcommand{\D}{{\bf D}}
\newcommand{\Kp}{{\bf K}'}

\newcommand{\ep}{\epsilon}
\newcommand{\be}{\begin{equation}}
\newcommand{\ee}{\end{equation}}
\newcommand{\T}{{\bf T}}
\newcommand{\G}{{\bf \Gamma}}
\newcommand{\Gtilde}{{\bf \tilde{\Gamma}}}
\newcommand{\C}{{\bf C}}
\renewcommand{\a}{{\bf a}}
\renewcommand{\k}{{\bf k}}
\newcommand{\e}{{\bf e}}
\renewcommand{\r}{{\bf r}}

\newcommand{\zak}{{\cal Z}}

\newcommand{\Energ}{\epsilon}

\newcommand{\vecteur}{{\bf g}}

\begin{document}
\title{The Zak phase and the existence of edge states in graphene}

\author{P. Delplace}
\affiliation{D\'epartement de Physique Th\'eorique, Universit\'e de Gen\`eve, CH-1211 Gen\`eve, Switzerland}
\author{D. Ullmo}
\affiliation{Laboratoire de Physique Th\'eorique et Mod\`eles Statistiques, CNRS UMR 8626, Univ.
 Paris-Sud,  91405 Orsay Cedex, France}
\author{G. Montambaux}
\affiliation{Laboratoire de Physique des Solides, CNRS UMR 8502,
Univ. Paris-Sud, 91405 Orsay Cedex, France}

\begin{abstract}
We develop a method to predict the existence of edge states in graphene ribbons for a large class of boundaries.
 This approach is based on the bulk-edge correspondence
between the quantized value of the Zak phase ${\cal Z}(k_\parallel)$, which is a Berry phase
across an appropriately chosen one dimensional Brillouin zone, and the existence of a localized
state of momentum $k_\parallel$ at the boundary of the ribbon. This bulk-edge correspondence is
rigorously demonstrated  for a one dimensional toy model as well as  for
graphene ribbons with zigzag edges. The range of $k_{\parallel}$ for which
edge states exist in a graphene ribbon is then calculated for arbitrary orientations of the edges.
Finally, we show that the
introduction of an anisotropy leads to a topological transition in terms of the Zak phase, which modifies the
localization properties at the edges. Our approach gives a new geometrical understanding of edge states, it
confirms and generalizes the results of several previous works.
\end{abstract}

\maketitle

\section{Introduction}

The physics of edge states in two-dimensional (2D) systems has
emerged as a very challenging problem  in solid state physics. A beautiful
illustration occurs in graphene, a monolayer crystal of
carbon,\cite{novoselov} where the existence of such states was
predicted\cite{fujita96,nakada96} in 1996 and confirmed
experimentally later in graphene\cite{kobayashi,niimi} and graphene-like structures.\cite{kuhl} This remarkable
feature led to a strong activity  during the
last few years. For instance, edge states were predicted to give rise
to a novel type of magnetic ordering\cite{fujita96} and may lead to
the realization of novel spintronic devices.\cite{louie,yazyev}

In a broader context, edge states are also known to play an important
role in quantum Hall systems\cite{halperin,buttiker} and topological
insulators.\cite{bernevig06,konig08} Because of
their chiral character, the edge states in quantum Hall systems are
robust against all kind of disorder or interactions, while those in
topological insulators survive scattering that preserves the time
reversal symmetry. This robustness against weak perturbations can be
understood from a correspondence between the number of edge states and
the value of a bulk topological number which is
basically the Berry curvature integrated over the space of parameters,
that is the Brillouin zone of the 2D
system\cite{TKNN,konig08}.

\begin{figure}[h!]
  \centering
  \includegraphics[width=7.6cm]{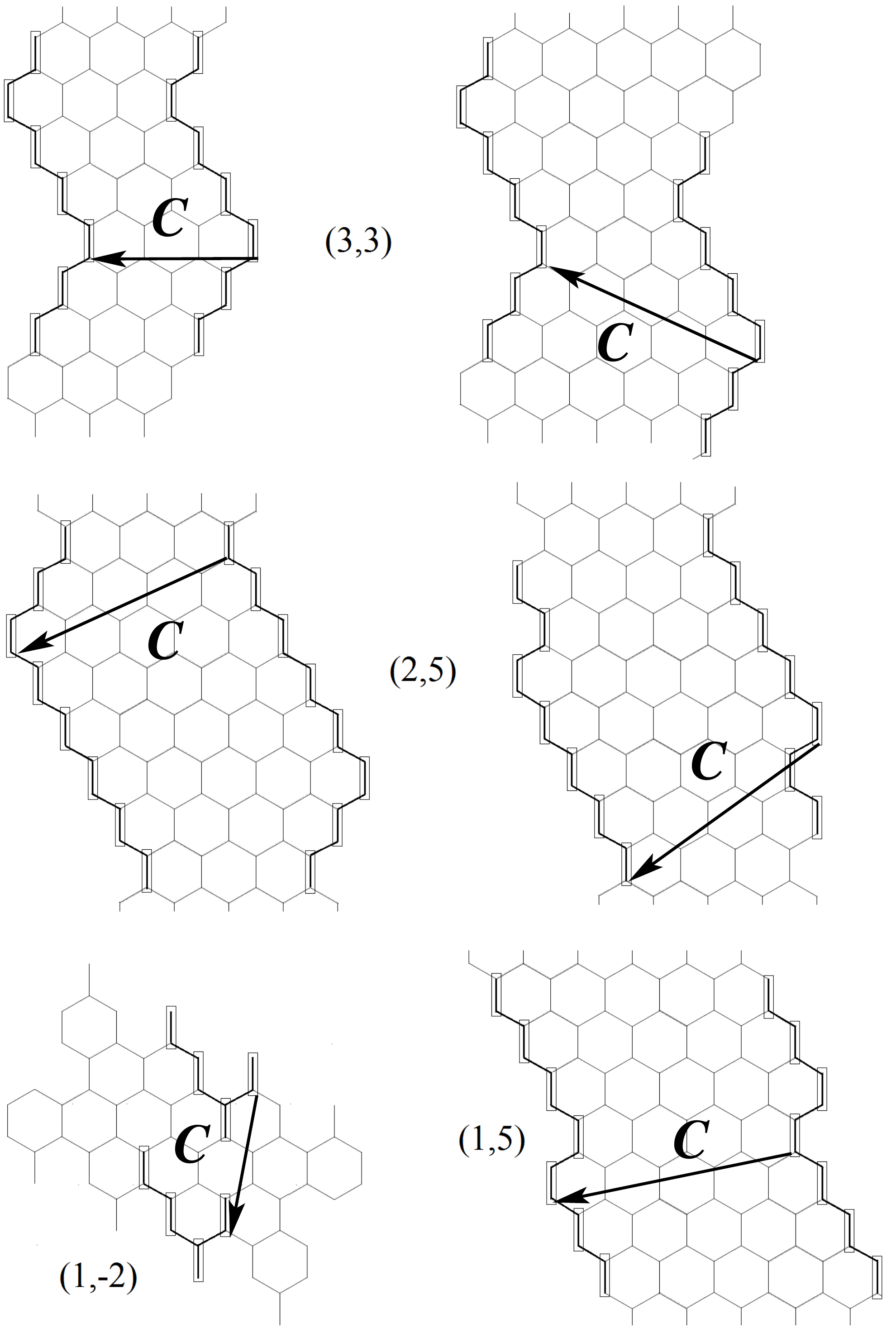}
  \caption{Examples of edges of different ribbons studied in this paper. They are characterized by a translation vector
    ${\T}(m,n)$, see section \ref{sec:edge-graphene-ribbons}. The values of the couples $(m,n)$ are specified in
    the figure. The edges of the ribbon $(1,-2)$ shows dangling bounds
    since $mn<0$. Different ribbons can be obtained from the same
    vector ${\T}(m,n)$ in two different ways. For instance, the lattice
    vectors $\C$ connecting the left and right edges of the two ribbons
    defined by ${\T}(3,3)$ are different.  An other possibility is to
    draw a different ``unit-cell'' pattern for the same vector $\T$,
    as for the two ribbons defined by ${\T}(2,5)$.}
		\label{fig:bords}
\end{figure}

Edge states in graphene differ from those mentioned above, the most
important distinction being that their existence depends on the
boundary conditions fixed by the shape of the
edge\cite{fujita96,nakada96}. Then, two questions naturally arise. The
first one is related to the bulk-edge correspondence:\cite{hatsugai02,mong10,sasakiNJP10,jianli_prb10} if the
localization  of a state at the edge  depends on the boundaries, is it
possible to relate its existence to a topological quantity defined within the
bulk ?  The second question is simply whether we can predict the
existence of edge states for an arbitrary type of edge (see e.g. some examples on Fig.~\ref{fig:bords}).

These two stimulating questions have already led to many works. In
particular, Ryu and Hatsugai  showed that
edge states in 2D systems with chiral symmetry can be related to a
bulk topological number defined in a reduced (1D) space of
parameters.\cite{hatsugai02} More precisely, these authors were
able to characterize edge states for three simple types of boundaries, namely the
zigzag, armchair and bearded edges (a bearded edge is a zigzag edge with dangling bonds, also called Klein defects \cite{klein99,liu09}). Of course, even without
disorder, there is an infinite number of different
edge geometries in a honeycomb lattice, and several recent theoretical
works addressed the existence of edge states for  more
sophisticated boundaries. Several of these works consist in
tight-binding calculations of ribbons band
structures.\cite{fujita96,nakada96,wakabayashi09,wakabayashi_jap10,jaskolski11}
 A general study including many various shapes of edges
  was also achieved by Akhmerov and Beenakker.\cite{akhmerov} This
  work, performed within the continuous (Dirac) approximation, notably
  provides an analytical formula for the density of edge states.

  The aim of this work is to precise the bulk-edge correspondence in
  graphene and to address a new method to predict the existence of
  edge states for  a large class of edges.  We
  show that it is possible to define in an unambiguous way a
  topological phase from the bulk Hamiltonian of graphene, which
  properly takes into account the shape of the edges.  For a 1D
  system, this phase, called Zak phase, is nothing but the integration
  of the Berry connection over the first Brillouin zone\cite{zak}
\begin{equation}
{\cal Z} = i\oint dq \left\langle u_{q}|\partial_{q}u_{q}\right\rangle
\; ,
\label{zakdef}
\end{equation}
where the $\left|u_q\right\rangle$ are the Bloch wave functions. In a
2D system, one  difficulty is to define properly the path over
which the integration is performed, and to relate this quantity to the
nature of the edge. In particular, for a translation invariant system
in one direction, the Zak phase depends on the crystal momentum
$k_{\parallel}$ associated with this direction. Here, we show that
${\cal Z}(k_\parallel)/\pi$ gives the number of states localized at
the edge of the system.

The outline of the paper is as follows. As the Zak phase is defined
 as a one-dimensional integral of the Berry connection, we first focus in
Sec. II on a one-dimensional toy model for a chain of dimers. For
this case, we give a simple demonstration of the bulk-edge
correspondence between the Zak phase and the existence of edge states.  Next, in
Sec. III, we turn to graphene where a similar demonstration is
performed for zigzag edges. By a formal analogy with the chain of
dimers, we assume the generalization of this bulk-edge correspondence
for the other boundaries. Then, we show how to define in an
unambiguous way the Zak phase in graphene
according to the nature of the edge and propose a very simple
graphical method to evaluate it. This information
then directly gives us the range of $k_\parallel$ for which edge
states exist. The analytical results are in perfect
agreement with those of Akhmerov and Beenakker \cite{akhmerov} and
reproduce many numerical works.  Finally, in Sec IV we extend our
approach by considering non-equal hopping parameters in the honeycomb
lattice. We discuss the existence of edge states in this case
 and explain recent numerical calculations in terms of the Zak
phase.\cite{dahal10}

\section{The Zak phase and the edge states in the chain of dimers}
\label{sec:chain}

\begin{figure}[htb]
\centering
\includegraphics[width=9cm]{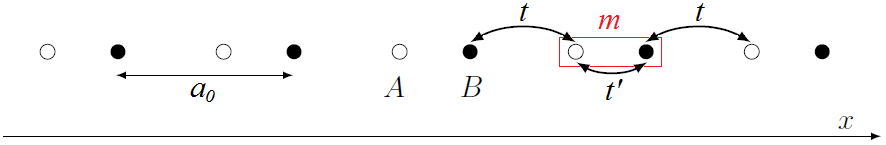}
\caption{Chain of dimers $A$-$B$. The chain starts with an atom $A$ and
  ends with an atom $B$. $t$ and $t'$ are the hopping parameters and $a_0$ is the lattice spacing. The
  unit cell $m$ is represented by a rectangle.}
	\label{fig:chaine}
\end{figure}

To illustrate the relation between the Zak phase and boundary states,
we consider in this section a simple model, a one-dimensional chain of
dimers $A$-$B$ as shown in Fig.~\ref{fig:chaine}. The two atoms of
the dimers are coupled by a hopping parameter $t'$ and the chain is
obtained by coupling periodically the dimers with a hopping parameter
$t$. The lattice spacing (between two consecutive identical atoms) is
$a_0$.

The aim of this section is to show in a simple way that the Zak phase
$\zak$ is governed by the ratio $t'/t$, and that the topological
transition $ (\zak \!= \!0) \rightarrow (\zak \!= \! \pi)$ corresponds
to the emergence of edge states in the finite system.

\subsection{The bulk Hamiltonian}

The Hamiltonian of the dimer chain is given by
\begin{equation}
\hat H = \sum_{m=1}^M t' b_m^\dagger a_m + t a_{m+1}^\dagger b_m + h .c. \; ,
\label{tightbinding}
\end{equation}
where $a_m^\dagger$ (resp.\ $b_m^\dagger$) creates  a particle
on the site $A$ (resp.\ $B$) of the $m^{th}$ dimer.

For periodic boundary conditions, we can use the
Bloch theorem and rewrite $\hat H$  as
\begin{equation}
\hat H = \sum_{k_n \equiv \frac{2 \pi n}{a_0}} \mathbf{\Psi}^\dagger_{k_n} \mathcal{H}^B(k_n)
\mathbf{\Psi}_{k_n}  \qquad \mbox{$(-\frac{M}{2} <n<\frac{M}{2}) $}\; ,
\end{equation}
with
 $\mathbf{\Psi}^\dagger_k \!=\! (\psi^\dagger_{A,k}, \psi^\dagger_{B,k})
\!=\! M^{-1/2} \sum_{m=1}^M  e^{i a_0  m k}( a_m^\dagger,b_m^\dagger)$ and
\begin{equation}
\mathcal{H}^B(k)=-t  \left(
\begin{array}{cc}
0&\rho(k)\\
\rho^*(k)&0\\
\end{array} \right)\ ,
\label{tbchain}
\end{equation}
where $\rho(k)=t'/t+e^{-ika_0}$.
Introducing $\mathbf{\sigma}=(\sigma_x,\sigma_y)$ the vector of Pauli
matrices, $\mathcal{H}^B(k)$ can be expressed in the form
\begin{equation}
\mathcal{H}^B(k)=-t \, \vecteur(k) \cdot \mathbf{\sigma}
\label{bulkgchain}
\end{equation}
with $\vecteur(k) = (\mathbbm{R}e\ \rho, -\mathbbm{I}m\ \rho )=(t'/t + \cos k a_0, \sin k a_0 )$.
Diagonalizing  $\mathcal{H}^B(k)$  we obtain the eigenvalues
\begin{equation}
\begin{split}
 \Energ_{k,\pm} &=\pm t\ |\vecteur(k)|= \pm t\ |\rho(k)|\\
  &=\pm \sqrt{t^2 + t'^2 +
   2tt'\cos(ka_0)} \; ,
 \end{split}
    \label{spectre}
\end{equation}
and writing $\rho(k)=|\rho(k)|e^{-i\phi(k)}$, we have
\begin{equation}
\vecteur(k) = |\rho(k)|
 \left(
\begin{array}{c}
\cos \phi(k)\\
\sin \phi(k)
\end{array}
\right) \ ,
\label{hamgchain}
\end{equation}
with the phase  $\phi(k)$ given by
\begin{equation}
\cot\phi(k)= \frac{t'/t}{\sin ka_0}+\cot ka_0 \; .
\label{eq:cot}
\end{equation}

\begin{figure}[htb]
\centering
\includegraphics[width=9cm]{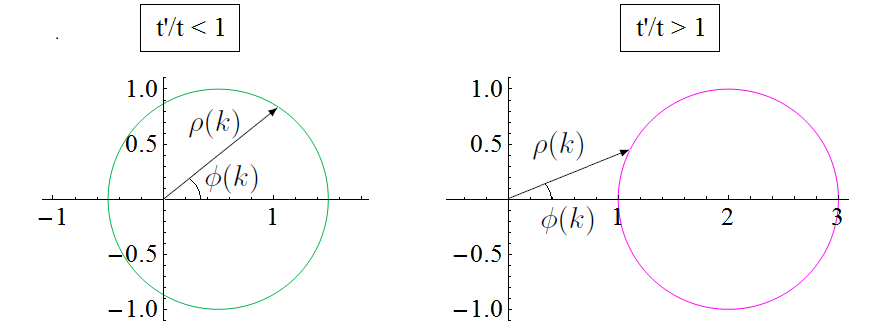}
\caption{Two trajectories of the vector $\vecteur(k)$ with different
  topologies when $k$ runs across the Brillouin zone. When
  $t'/t<1$ ($t'/t>1$) the curve $\vecteur(k)$ does (not) wind around the
  origin.}
\label{fig:cercleh}
\end{figure}

The winding of the vector $\vecteur(k)$ as $k$ varies across the
Brillouin zone is shown in Fig.~\ref{fig:cercleh} for two values of
$t'/t$. When $t'/t>1$, the curve $\vecteur(k)$ does not enclose the
origin and $|\phi(k)|<\pi/2$ for all $k$. When $t'/t<1$, the loop
encloses the origin and the phase $\phi(k)$ can take any value. This  topological behavior of the
phase $\phi(k)$ is furthermore closely related to the value of the Zak phase.
Indeed, the eigenvectors of $\mathcal{H}^B(k)$ are of the form
\begin{equation}
\left|u_{k,\pm}\right\rangle =
\frac{1}{\sqrt{2}}
\left(
\begin{array}{c}
e^{-i\phi(k)}\\
\pm 1
\end{array}
\right) \; ,
\label{wfbulkchain}
\end{equation}
and the definition Eq.~(\ref{zakdef}) of the Zak phase gives
\begin{equation} \label{eq:zakexplicit}
\zak = \frac{1}{2} \oint dk \frac{d \phi}{d k} =
\frac{\Delta\phi}{2} \; ,
\end{equation}
where $\Delta\phi$ is the variation of  $\phi(k)$ when $k$ varies across
the full Brillouin zone.  The Zak phase $\zak$ is
$\pi$ times the winding number of the curve $\vecteur(k)$ around the
origin, and is therefore zero if this curve does not enclose the
origin and $\pi$ if it does.  Thus
\begin{equation}
\begin{split}
{\cal Z}&=0 \qquad  \text{when}\ t'/t>1 \\
{\cal Z}&=\pi \qquad  \text{when} \ t'/t<1 \ .
\end{split}
\label{tuning}
\end{equation}
 Eq.~(\ref{tuning}) shows that
tuning the ratio $t'/t$ induces a topological transition (at $t'/t\!=\!1$)
characterized by the Zak phase ${\cal Z}=0 \longleftrightarrow {\cal
  Z}=\pi$.

\subsection{Open boundary conditions}
\label{sec:open}
\subsubsection{The missing bulk states and the Zak phase}

\begin{figure}[htb]
\includegraphics[width=8cm]{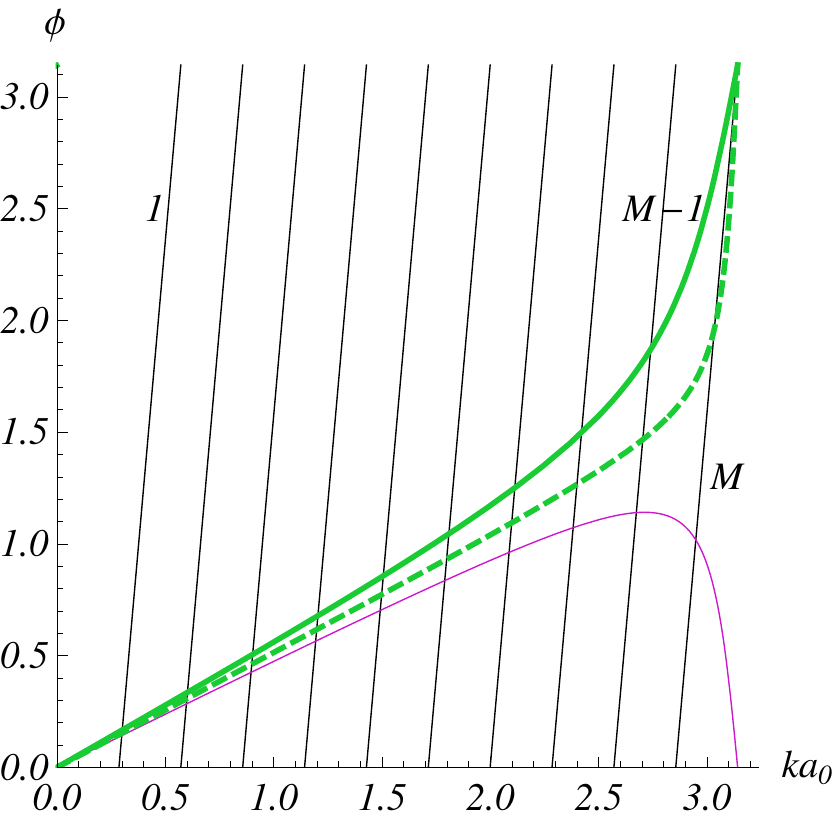}
 \caption{ Variation $\phi(k)$ for $t'/t=1.1$ (bottom curve), $t'/t=0.8$
   (upper thick curve), and $t'/t=0.95$ (thick dashed curve). The straight lines
   are lines of equation $f_{\kappa} (k)= (M+1) k a_0 - \kappa \pi$. Here, we
   have chosen $M=10$, for which $(t'/t)_c= 0.9091$. The extreme
   values of $\kappa$ are indicated on the figure. In the latter case
   $t'/t=0.95 > (t'/t)_c$, so that there are $M$ bulk states although
   the Zak phase $\zak=\pi$. }
\label{fig.phidek}
\end{figure}

Consider now a finite chain of $M$ dimers with open boundary
conditions. We impose that the wave function vanishes at
the nearest site outside the chain, that is on the $B$ site at
$m=0$ and the $A$ site at $m=(M+1)$.  Most of the eigenvectors can be
constructed as linear combinations $|v_{k,\pm}\rangle$  of the bulk
eigenfunctions with opposite momentum $|u_{k,\pm}\rangle$ and
$|u_{-k,\pm}\rangle$.   Writing
$|u_{k,\pm}\rangle$ as
\begin{equation}
\left|u_{k,\pm}\right\rangle = \sqrt{\frac{1}{2M}}
\sum_{m=1}^M{e^{ik a_0 m}}
\left(
\begin{array}{c}
e^{-i\phi(k)}\\
\pm 1
\end{array}
\right)\cdot \left(\left|m,A\right\rangle ,\left|m,B\right\rangle
\right) \; ,
\label{wfchain}
\end{equation}
where $\left|m,{A/B}\right\rangle$ denotes the orbital $A$/$B$ in the
cell $m$, we obtain from the condition $ \langle 0,B |v_{k,\pm}\rangle
=0 $ that $|v_{k,\pm}\rangle = \frac{1}{\sqrt{2}} \left[ |u_{k,\pm}
  \rangle - \left|u_{-k,\pm}\right\rangle \right]$. Thus, using
  $\phi(-k)=-\phi(k)$, we have
\begin{equation} \label{eq:Vk}
\left|v_{k,\pm}\right\rangle = \frac{i}{\sqrt{M}}
\sum_m{ \left(
    \begin{array}{c}
      \sin(k a_0 m - \phi(k))\\
      \pm \sin(k a_0 m)
    \end{array}  \right)}
\cdot \left(\left|m,A\right\rangle ,\left|m,B\right\rangle
\right) \; .
\end{equation}
Finally, the boundary condition   $  \langle (M+1),A
|v_{k,\pm}\rangle =0 $ imposes the quantization condition
  \begin{equation}
k(M+1) a_0 - \phi(k) =\kappa \pi , \qquad \kappa=\ 1,\ \cdots M
 \label{eq:qc} \; ,
\end{equation}
which has to be solved in the range $0<k<\pi/a_0$ (the wave
  functions corresponding to $k=0$ and $k=\pi/a_0$ are identically zero).
 The function $\phi(k)$ is plotted in
Fig.~\ref{fig.phidek}, and the solutions of Eq.~(\ref{eq:qc})
correspond to the intersection of $\phi(k)$ with the $M$ lines $f_{\kappa}(k)= (M+1) k a_0 - \kappa \pi$.

From this figure, we see that the Zak phase controls the number of bulk states and
therefore the existence of edge states. Indeed,
when $t' > t$, $\phi(\pi/a_0)=0$ and Eq.~(\ref{eq:qc})
has $M$ solutions.
When $t' < t$, $\phi(\pi/a_0)=\pi$ and in this case there are $M$
or $M-1$ solutions depending on the value of $t'/t$. By comparing the
slopes of the curves $\phi(k)$ and $f_M(k)$, we immediately obtain that
the number of bulk states $\left|v_{k,\pm}\right\rangle$ depends
  on the critical value of the ratio $t'/t$
 \begin{equation}
\left( \frac{t'}{t} \right)_c= 1 - \frac{1}{M+1} \,  .
\label{eq:ratc}
\end{equation}
Including the factor 2 associated with  negative and
 positive energies for each solution of the quantization condition
(\ref{eq:qc}), the number $N_{\rm bulk}$ of bulk states is:
\begin{equation}
\begin{split}
&N_{\rm bulk}=2M \qquad \qquad \ \ \text{when} \ t'/t > \left( {t' /
    t} \right)_c \\
&N_{\rm bulk}=2(M-1) \qquad \text{when} \ t'/t < \left( {t' / t} \right)_c \; .
\end{split}
\label{critic}
\end{equation}
As we show below, the missing states  are edge states localized at the ends
of the chain.
In the large $M$ limit, the number of bulk states is related to the
value of $\phi(\pi/a_0)$ since, in this limit, there are $2 M$
bulk solutions when $\phi(\pi/a)=0$ and
$2(M-1)$ bulk solutions when $\phi(\pi/a)=\pi$.
This criterion can be rewritten in terms of the Zak
phase: since $\phi(k)$ is an odd function of $k$, we have simply
\begin{equation}
{\cal Z}=\frac{1}{2} \oint dk \frac{d \phi}{d k }= \int_0^{\pi/a_0} dk
\frac{d \phi }{ d k }= \phi(\pi/a_0) = 0 \ \mbox{or} \ \pi
\end{equation}
As a result, by comparing Eqs.~(\ref{tuning}), (\ref{eq:ratc}) and
  (\ref{critic}), we conclude that in the large $M$
limit,\footnote{When $M$ is finite, there is a finite range of
  parameters $1 - 1/(M+1) < t'/t <1$, for which there are no edge
  states ($M$ bulk states), although the Zak phase is $\pi$.}
\begin{equation}
\begin{split}
 N_{\rm bulk} &= 2M    \qquad \qquad \ \  \text{when} \  {\cal Z} =0 \\
 N_{\rm bulk} &= 2(M-1) \qquad  \text{when} \  {\cal Z}=\pi \; .
\end{split}
\label{eq:Nbulk}
\end{equation}


\subsubsection{The edge states}

\begin{figure}[htb]
\includegraphics[width=7cm]{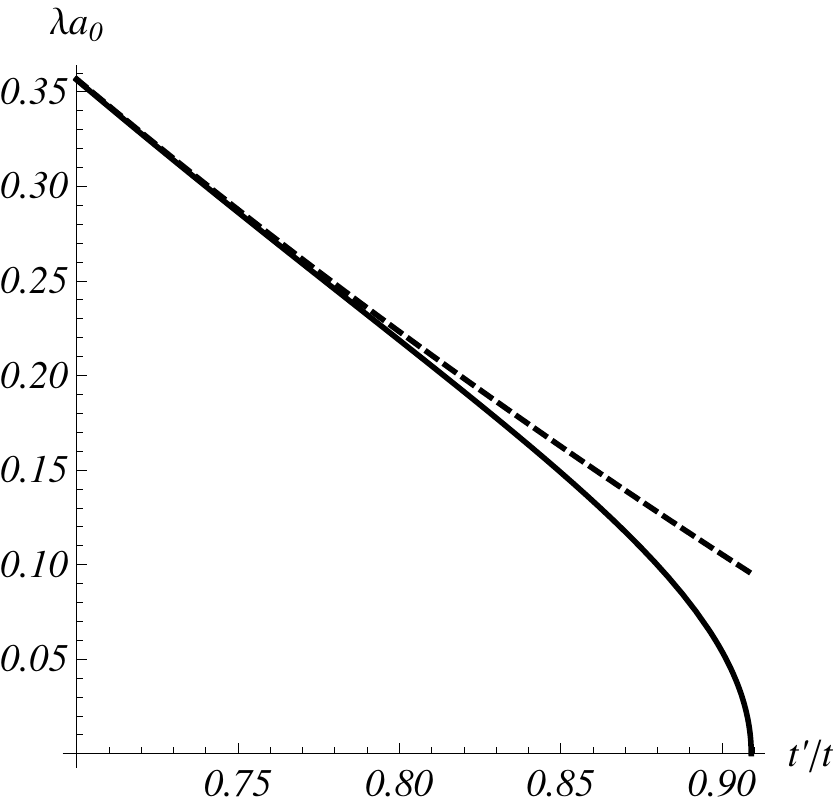}
\caption{ Inverse localization length $\lambda$ as a function of the
  parameter $t'/t$, for a chain of $M=10$ dimers. The full curve is
  the solution of Eq.~(\ref{qcedge}). As expected, it diverges as
    the ratio $t/t'$ reaches the critical value given by
    Eq.~(\ref{eq:ratc}). The dashed line corresponds to the
  approximation (\ref{lambdaapprox}) valid far from the transition and
  corresponding to $\Energ=0$. }
\label{fig.lambda}
\end{figure}

\begin{figure}[htb]
\includegraphics[width=8cm]{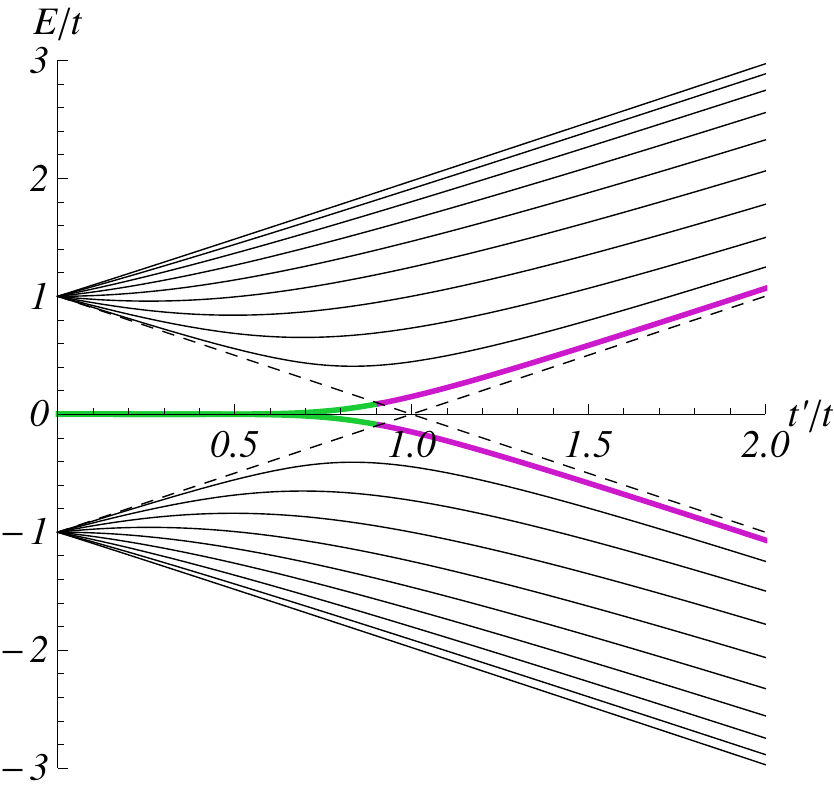}
 \caption{Energy levels as a function of the parameter $t'/t$, for a
   chain of $M=10$ dimers. There is an edge state with an energy close to $0$ when $t'/t <
   1 -1/(1+M)$ . The dashed lines correspond to the gap $\pm (t'-t)$ in
   the limit $M \rightarrow \infty$. The thick (dark, purple online) curve is the energy of the
   lowest energy state which becomes an edge state (light, green online) when
   $(t'/t)_c < 1$.}
\label{fig:Etpt}
\end{figure}

  We now briefly describe the structure of the edge state  for $t'/t <
  (t'/t)_c$. We search for a solution $k$ of the form  $k   =
  \pi/a_0 + i  \lambda$, where $\xi=1/\lambda$ is the localization
  length of the edge state.
 The solution vanishing on the $B$ site at $m=0$ is of the form
  \begin{equation}
 \left|v^e_{\lambda,\pm}\right\rangle = \frac{1}{\sqrt{M}} \sum_{m=1}^M{(-1)^{m+1}}
  \left(
    \begin{array}{c}
      \chi^A_m   \\
      \chi^B_m
    \end{array}  \right)
    \cdot \left(\left|m,A\right\rangle ,\left|m,B\right\rangle
\right) \; ,
 \end{equation}
 where
  \begin{equation}
 \left(
    \begin{array}{c}
      \chi^A_m   \\
      \chi^B_m
    \end{array}  \right) = \left(
                  \begin{array}{c}
                    \left(\frac{t'}{t} \sinh  \lambda  a_0 m -   \sinh  \lambda a_0 (m-1)\right)/{|\rho(\lambda)|} \\
                     \pm\sinh  \lambda  a_0 m  \\
                  \end{array}
                \right) \; ,
               \label{edgestate1}
\end{equation}
and has an energy
 \begin{equation}
 \Energ_{\lambda,\pm} = \pm\sqrt{t^2 + t'^2 - 2 t t' \cosh \lambda a_0 } \equiv \pm t\
 |\rho(\lambda)|
 \; .
\label{energielambda}\end{equation}
The inverse localization length $\lambda$ is fixed by the condition that the wave
function on the site $A$ at $m=(M+1)$ vanishes, leading to
\begin{equation}
 t' \sinh \lambda (M+1) a_0 = t \sinh \lambda M a_0  \label{qcedge} \;  .
\end{equation}
Inserting (\ref{qcedge}) into (\ref{energielambda}), we find 
\be \ep_{\lambda,\pm} = \pm t  \frac{\sinh \lambda a_0}{\sinh \lambda (M+1) a_0} \; , \label{enlambda} \ee  
and the components of the edge states wave functions can be rewritten as
  \begin{equation}
 \left(
    \begin{array}{c}
      \chi^A_m   \\
      \chi^B_m
    \end{array}  \right) = \left(
                  \begin{array}{c}
                    \sinh \lambda a_0 (M+1-m) \\
                     \pm \sinh  \lambda  a_0 m  \\
                  \end{array}
                \right) \; ,
              \end{equation}
which satisfy properly the boundary conditions.
Far from the transition, that is when the localization length
  $\xi$ is much smaller than the size of the chain $M a_0$,  Eq.~(\ref{qcedge}) reads
\begin{equation}
 \frac{t'}{t}  \simeq \exp(-\lambda a_0)  \label{lambdaapprox} \; .
\end{equation}
This   implies $ \cosh(\lambda a_0) \simeq (t^2 + t'^2) /2 t t'$,
and thus $\Energ_{\lambda,\pm}  \simeq 0$.   More precisely, from Eq.~(\ref{enlambda}), the energy vanishes as $\Energ_{\lambda,\pm}  \simeq \exp(- \lambda M a_0)$. The dependence $\lambda(t'/t)$ for a
  chain of $M=10$ dimers is shown in Fig.~\ref{fig.lambda},
together with the approximate expression Eq.~(\ref{lambdaapprox}).
The energy spectrum of the same chain is displayed in Fig.~\ref{fig:Etpt}.

 \subsection{Remarks on the chiral symmetry}

 We finish this section with a few brief remarks concerning the
   Zak phase and symmetries.  As first stressed by Ryu and Hatsugai,
 \cite{hatsugai02} the fact that the edge states have zero energy is
 associated with the existence of a chiral symmetry of the bulk
 Hamiltonian.\footnote{The energy of the edge states is  zero
   only when the width of the system is larger than the localization
   length, otherwise the edge states at each edge hybridize and the
   resulting energy is not zero.}  From an algebraic point of view, a
 chiral symmetry is represented by an operator ${\cal C}$ which
 anticommutes with the bulk Hamiltonian and which satisfies ${\cal
   C}^2=\mathbbm{1}$. As the Bloch Hamiltonian of the chain of dimers
 can be written as a linear combination of the Pauli matrices
 $\sigma_x$ and $\sigma_y$, it is clear that ${\cal C}=\sigma_z$
 fulfills these properties. Now, we note that the Zak phase measures the
 solid angle drawn by the pseudo-spinor in the Bloch sphere when $k$
 spans the Brillouin zone. As long as the Hamiltonian does not have a
 component proportional to $\sigma_z$, the pseudo-spinor is forced to
 evolve on the equator of the Bloch sphere, and the Zak
 phase is necessarily a multiple of $\pi$. We notice that breaking the inversion symmetry of
 the chain,\cite{zak} for instance by adding a staggered potential, would add a term proportional to $\sigma_z$ in the
 Hamiltonian and therefore would break the chiral symmetry. As a
 consequence, the Zak phase is not expected to be a multiple of $\pi$
 anymore in this case and the energy of the edge states can be
 different from zero.

\begin{figure}[htb]
\centering
\includegraphics[width=9cm]{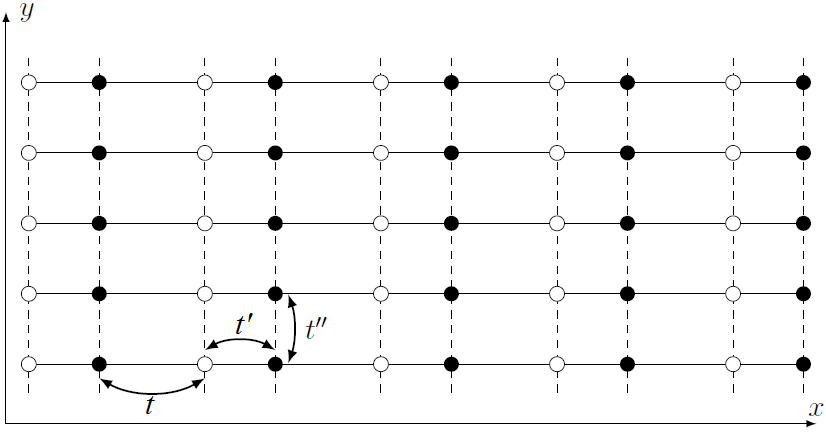}
\caption{Chains of dimers coupled in the $y$ direction by a
    hopping parameter $t''$.}
	\label{fig:chainescouplees}
\end{figure}

\begin{figure}[htb]
\centering
\includegraphics[width=4.3cm]{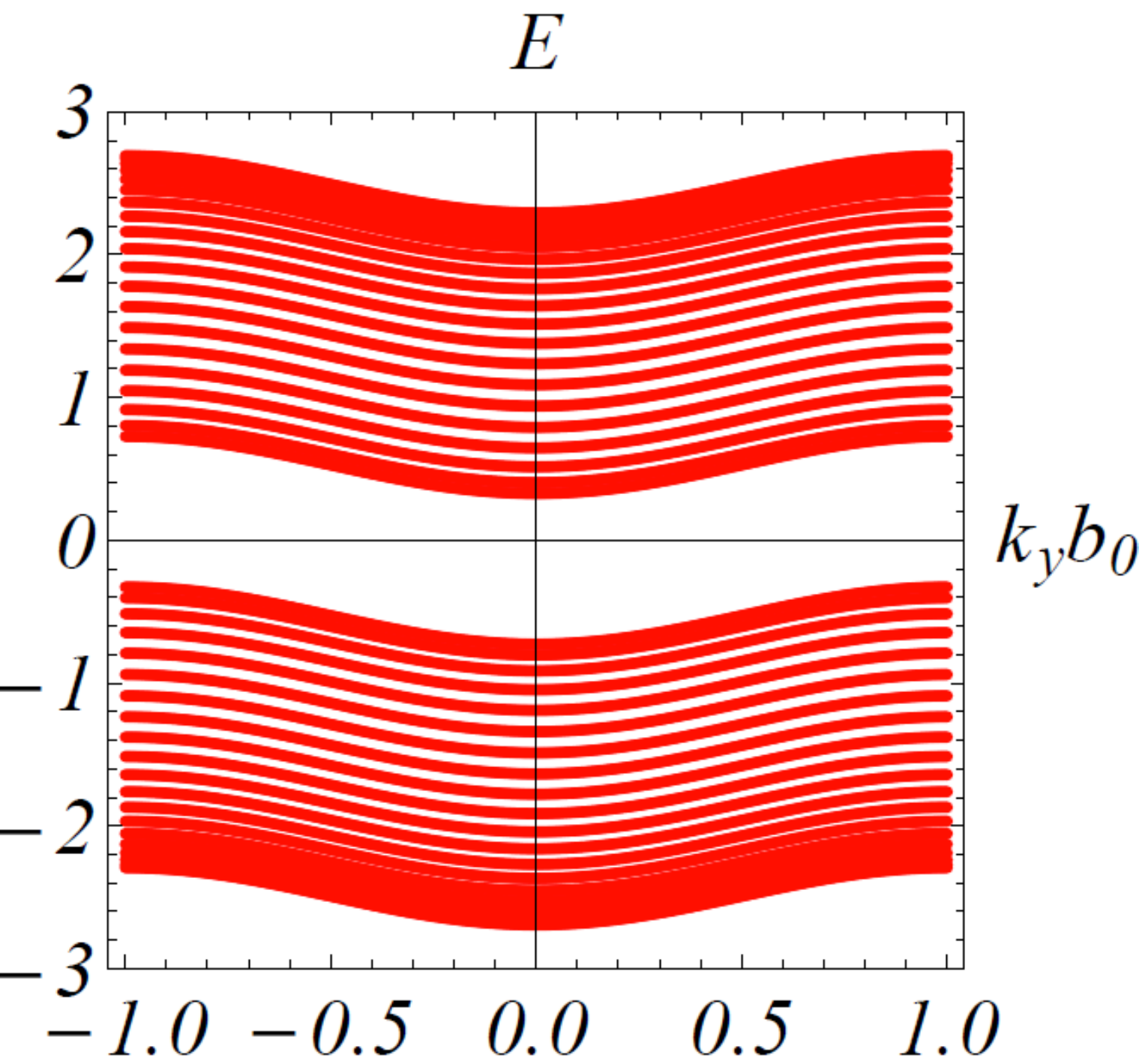}
\centering
\includegraphics[width=4.3cm]{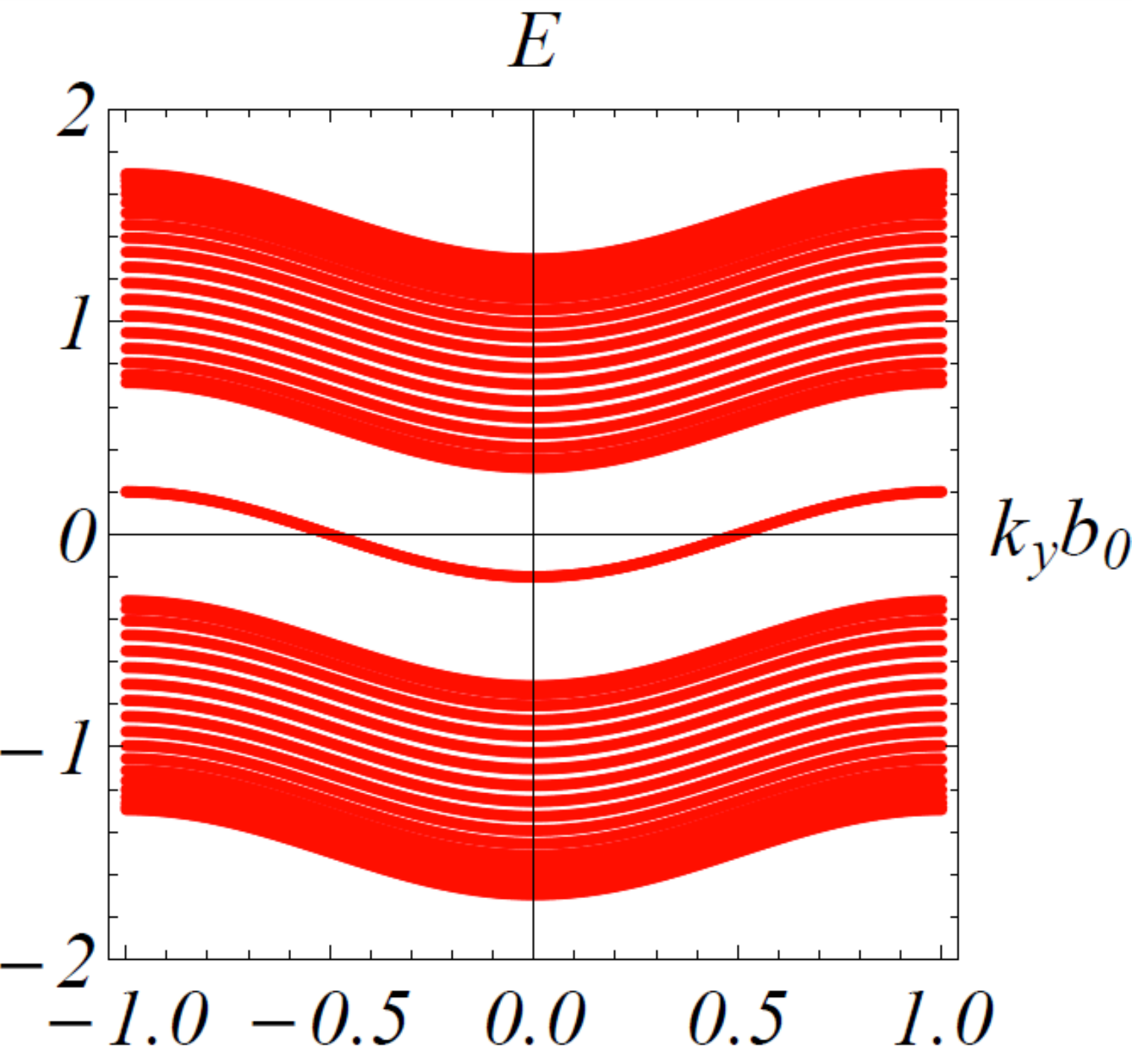}
\caption{Energy levels of a ribbon built from chains of $M=20$
  dimers. The chains are coupled one to each other in the $y$
  direction with a hopping parameter $t''=0.1$, and the ribbon is invariant by translation along this direction.
  We took $t=1$, and (left) $t'=1,5$, (right) $t'=0,5$. In the first
  case the Zak phase is ${\cal Z}=0$ and there is no edge state. In
  the other case, the Zak phase is ${\cal Z}=\pi$ and dispersive edge
  states have emerged in the gap.}
	\label{fig:rubans_peierls}
\end{figure}

 Adding a term proportional to the identity trivially breaks the
 chiral symmetry and shifts the energy while leaving the pseudo-spinor
 on the equator so that the Zak phase is still quantized as a multiple
 of $\pi$.    A configuration where this simple mechanism leads to an
 interesting physics is obtained by coupling
  chains of dimers (assumed oriented along the $x$-direction) by a hopping
  parameter $t''$ along the $y$ direction, as displayed in
  Fig.~\ref{fig:chainescouplees}.
For periodic boundary conditions, the bulk Hamiltonian of
the coupled chains reads
\begin{equation}
{\cal H}^B_{cc}=-t''\cos(k_yb_0)\mathbbm{1}+{\cal H}^B \; ,
\end{equation}
with $b_0$ the distance between two chains. We see that the term
proportional to $\mathbbm{1}$ involves a dispersion along the $y$
direction and does not change the properties of the Zak phase which
are encoded in ${\cal H}^B$. Therefore, two distinct topological phases
(${\cal Z}=0$ and ${\cal Z}=\pi$) arise when the criteria
given in Eq.~(\ref{critic}) is satisfied. This means that
when ${\cal Z}=\pi$, a ribbon of finite width in the $x$ direction and
invariant by translation in the $y$ direction supports dispersive edge
states along its edges, as shown in Fig~\ref{fig:rubans_peierls}.

This example shows that the Zak phase may help characterizing the edge
states even in the absence of a chiral symmetry. In addition, it also
illustrates that the Zak phase, which until now we have defined for a
one-dimensional system, may provide informations about the edge states in
systems of higher dimension. In the following of the paper, we
investigate the more complex case of a monolayer of graphene.

\section{The Zak phase and the edge states of  graphene ribbons}

Turning now to graphene, we demonstrate in this section that the Zak
phase, introduced in section \ref{sec:chain} for a one-dimensional
system, has a natural generalization for a large class of two-dimensional graphene ribbons.  In the particular case of zigzag
edges, following the same lines as in the 1D case, we prove that it
is possible to relate the Zak phase to the existence of edge
states.  We then consider the case of ribbons with arbitrary
orientations, and show that, computing the Zak phase, we can
predict the presence or absence of edge states according to the nature
of the edge.

\subsection{The bulk Hamiltonian}

\begin{figure}[ht]
  \centering
  \includegraphics[width=9cm]{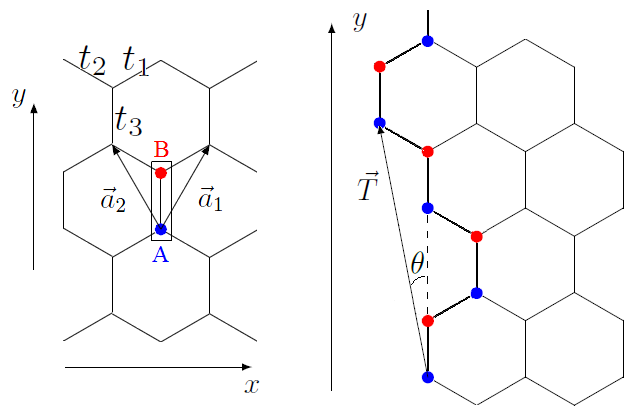}
  \caption{(Left) Unit cell (dimer) $A$-$B$ of the graphene sheet
    with the basis vectors of the Bravais
    lattice ${\a}_1$ and ${\a}_2$  and the hopping parameters $t_1$, $t_2$ and
    $t_3$. (Right) Example of an edge obtained by translating the dimer along ${\a}_1$, then twice along
${\a}_2$. This edge is characterized by the periodicity vector ${\T}={\a}_1+2{\a}_2$.}
	\label{fig:gph}
       \end{figure}

We describe the electronic spectrum of graphene by a tight-binding
model on the triangular Bravais lattice with two atoms ($A$ and $B$) per
unit cell, as illustrated in Fig. \ref{fig:gph}. The parameters $t_1$,
$t_2$ and $t_3$ represent the three hopping integrals between
nearest neighbors and for now we consider the isotropic case
$t_1=t_2=t_3$ (the anisotropic case is treated in
Sec. \ref{sec:asym}).  The vectors ${\a}_1$ and ${\a}_2$ form a basis
of the Bravais lattice, and we note $({\a}^*_1,{\a}^*_2)$ (with
${\a}^*_i \cdot {\a}_j = 2\pi \delta_{ij}$) the associated basis of
the reciprocal space.
 For periodic boundary conditions in both $x$ and $y$ directions,
  Bloch theorem leads to the bulk Hamiltonian
\begin{equation}
\begin{split}
H^B(\k)&=-t_3 |\rho(\k)| \left(
\begin{array}{cc}
0&e^{-i\phi(\k)}\\
e^{i\phi(\k)}&0\\
\end{array}
\right)\\
\end{split}
\label{eq:tb}
\end{equation}
in the basis of the two sub-lattices $A$ and $B$, with
\begin{equation}
\begin{split}
 \rho(\k)&=1+\frac{t_1}{t_3}\exp{(-i\k\cdot\a_1
   )}+\frac{t_2}{t_3}\exp{(-i\k\cdot\a_2 )}\\
   &=|\rho(\k)|e^{-i\phi(\k)} \; .
   \end{split}
 \label{eq:tb2} 
\end{equation}

The  eigenenergies of the bulk Hamiltonian $H^B(\k)$
consist in two bands,  given by $\ep_{\pm}(\k)=\pm
t_3|\rho(\k)|$, and   the
corresponding eigenvectors have the form,
\begin{equation}
  \left|u_{\k,\pm}\right\rangle = \frac{1}{\sqrt{2}}
  \left( \begin{array}{@{}c@{}}
          e^{-i\phi(\k)}\\
          \pm 1\\
          \end{array} \right) \ .
\label{eq:blochphase2d}
\end{equation}
The positive and negative bands  touch linearly at  two inequivalent
points $\D$ and ${\D}^{\prime}$ (the Dirac points) which, in
the isotropic case considered in this section, are located at the
corners $\K$ and $\Kp$ of the first Brillouin zone.

The Hamiltonian (\ref{eq:tb}) can be written in the same form
$H^B(\k)=-t_3\ \vecteur(\k)\cdot{\sigma}$ as for the chain of dimers
(\ref{bulkgchain}), the difference being that $\k$ is now a two-dimensional vector and that the $\k$ dependence of $\vecteur(\k)$ is
of course different.  As in our 1D toy model, topological properties
of the wave function as well as some characterization of the edge
states are expected to be encoded in the loops drawn by $\vecteur(\k)$
as $\k$ varies across the Brillouin zone. This connection was actually already suggested by
Ryu and Hatsugai \cite{hatsugai02} in the broader context of systems
with chiral symmetry (and more recently by Mong and Shivamoggi in a general study of Dirac Hamiltonians\cite{mong10}).
Their approach, which basically consists in the
graphical evaluation of the Zak phase in the same way as in
Fig. \ref{fig:cercleh}, was applied for three different regular types
of edges (zigzag, armchair and bearded edges). Comparing with
numerical calculations, they showed that the Zak phase could correctly
predict the existence of edge states in these cases. However, their approach
 relied on the construction of a \textit{specific}  bulk
  Hamiltonian (i.e. a vector $\vecteur(\k)$) for each type of edge
considered, and is therefore not convenient to consider arbitrary
boundary conditions.    Here, we keep the \textit{same} bulk Hamiltonian
but we associate to each type of edge  a \textit{specific}
   2D Brillouin zone. This allows us to
predict the existence of edge states for different ribbon geometries, and
therefore to address a significantly larger class of ribbons.

\subsection{Edges of graphene ribbons}

\label{sec:edge-graphene-ribbons}

Turning now to graphene ribbons, we assume that both edges of the
ribbon are parallel (i.e.\ that one edge can be deduced from the
  other by translation of a lattice vector $\C$) and constructed as
illustrated in Fig.~\ref{fig:gph}, i.e. \ by connecting dimers $A$-$B$ (or
unit cells) {\it of fixed orientation} (vertical orientation in
Fig.~\ref{fig:gph}).

More precisely, considering two positive or negative integers $m$
  and $n$, an edge is built as $(|m|+|n|)$ translations of the dimer,
  $|m|$ of which along ${\a}_1$ ($-{\a}_1$ if $m$ negative) and
  $|n|$ of which along ${\a}_2$ ($-{\a}_2$ if $n$ negative), in
  an arbitrary order, and by repeating the pattern obtained in this
  way. Therefore the edges are invariant under the  translation vector ${\T}=m {\a}_1 + n
{\a}_2$  which  characterizes the type of edge. Noting $\theta$ the angle
    between ${\T}$ and the $y$ axis, this angle is related to
    $(m,n)$ through
\begin{equation}
   \tan\theta=\frac{1}{\sqrt{3}}\frac{n-m}{n+m} \; .
   \label{theta}
\end{equation}
When $m$ and $n$ have the same sign, the class of edges constructed in
this way exactly corresponds to the minimal boundary conditions of
Akhmerov and Beenakker.\cite{akhmerov} In the other case, the
  edges exhibit dangling bonds.

Fig.~\ref{fig:bords} gives some examples of such edges. It is easy
to see that the vectors ${\T}(m,n)$ and ${\T}(n,m)$ describe the
same kind of edge.   We also notice that the
  same vector $\T$ can describe edges with different shapes, see
the example of ${\T}=(2,5)$ in Fig. \ref{fig:bords}.

We stress that choosing the ``unit-cell'' dimer $A$-$B$ with a
different orientation (i.e.\ rotated from $\pm 2\pi/3$ with respect to
the vertical one used in Fig.~\ref{fig:gph}) leads to a different set
of boundaries.  These latter are of course just deduced from the
former by a $\pm 2\pi/3$ rotation, and thus considering only the
vertical unit-cell dimer, as we shall do in the following, does not
restrict the type of edge to be studied.  We insist however that if
one wanted to consider another orientation of the unit-cell dimer for
the edges, it would be essential to \textit{modify accordingly the  dimer
  orientation} in the definition of the \textit{bulk} Hamiltonian
(which basically fixes the zero of the phase  $\phi(\k)$  in
Eqs.~(\ref{eq:tb}) or (\ref{eq:blochphase2d})). This is a necessary
condition to derive a bulk-edge correspondence for the edge states in
terms of the Zak phase.


\subsection{The Zak phase in graphene ribbons}

For one-dimensional models such as the one considered in
section~\ref{sec:chain}, the Zak phase is defined as the integral of
the Berry connection across the Brillouin zone.  To generalize this
notion to two-dimensional systems such as graphene, this integration
should be taken on a cut of a 2D Brillouin zone in a direction
transverse to the ribbon orientation. More precisely, as the ribbon is
assumed to be invariant under translation by the vector $\T$, Bloch theorem
guarantees that $k_\parallel$, the component of the crystal
  momentum parallel to $\T$, is a good quantum number.  We expect
therefore the Zak phase $\zak(k_\parallel)$ to be a function of
$k_\parallel$, and to correspond to an integration of the Berry
connection across the 2D Brillouin zone along a perpendicular
direction $k_\perp$.

Let us assume for now that $m$ and $n$ are coprime integers (we will
return below to the case where they are not).  We choose the Brillouin
zone from which the Zak phase will be computed as the one generated by
two orthogonal vectors of the reciprocal space, ${\G}_\parallel$ and
${\G}_\perp$, obtained as follows.  The first of these vector
${\G}_\parallel \equiv 2 \pi {\T} /|{\T}|^2$ is parallel to the
direction $\T$ of the ribbon and merely defines the one-dimensional
Brillouin zone of the ribbon.  The second vector ${\G}_\perp$ is taken
perpendicular to $\T$, and its norm is fixed by the constraint that
$\G_\parallel \times {\G_\perp}= {\a}_1^* \times {\a}_2^*$ (where, as
mentioned above, ${\a}_1^*$ and ${\a}_2^*$ are the reciprocal lattice
vectors defined by ${\a}_i \cdot {\a}_j^*=2 \pi \delta_{ij}$). This
leads to
\begin{eqnarray}
&\G_\parallel(m,n)&=\frac{(n+2m) {\a}_1^\ast +
(m+2n){\a}_2^\ast }{2(n^2+m^2+nm)}
\label{gammaparall}
\\
&\G_\perp(m,n)&=  n {\a}_1^\ast - m {\a}_2^\ast\; .
\label{gammaperp}
\label{Gammas}
\end{eqnarray}
For the sake of completeness we remind briefly in appendix~\ref{app:BZ}
why $({\G}_\parallel,{\G}_\perp)$  constructed in this way actually
defines a Brillouin zone when $(m,n)$ are coprime integers, and only in this case.

For an arbitrary edge characterized by the vector ${\T}(m,n)$, we
introduce the unit vectors $\e_\parallel=\G_\parallel /
|\G_\parallel|$ and $\e_\perp=\G_\perp /|\G_\perp|$, and write  the
momentum as $\k = k_\parallel \e_\parallel + k_\perp \e_\perp$. The Zak phase
$\zak(k_\parallel)$ can thus be defined as:
\begin{equation}
\zak (k_\parallel)=i\oint dk_\perp \left\langle
  u_{\k,\pm}|\partial_{k_\perp}u_{\k,\pm}\right\rangle \; ,
\end{equation}
 which, by using the expression of the Bloch function
(\ref{eq:blochphase2d}), simply reads
\begin{equation} \label{eq:zakII}
\zak (k_\parallel)=\frac{1}{2}\oint dk_\perp \partial_{k_\perp} \phi(\k)  \; .
\end{equation}
Note that as $\G_\perp$ is a vector of the reciprocal
lattice, the integration can indeed be seen as taken on a closed path.

Let us consider now the situation where $m$ and $n$ are not coprime,
in which case  $(\G_{\parallel},\G_\perp)$
obtained from Eqs.~(\ref{gammaparall}) and (\ref{gammaperp}) do not
form a basis of the reciprocal lattice (see appendix~\ref{app:BZ}).  Writing $m=l \tilde{m}$ and
$n= l \tilde{n}$ with $l$ integer and $\tilde m$ and $\tilde{n}$
coprime and following exactly the same line of argument
as above, we can construct a basis $( \Gtilde_\parallel,
\Gtilde_\perp)$ of the reciprocal lattice corresponding to ($\tilde
m,\tilde n$), and define the Zak phase $\zak_{(\tilde m,\tilde
  n)}(\tilde k_\parallel)$ accordingly from (\ref{eq:zakII}).  An
example of such a construction will be shown in Sec.~\ref{sec:general}
(see the case $(m,n) = (2,0)$ of Fig.~\ref{fig:bz}).  However, as
${\T}(\tilde{m},\tilde{n})={\T}(m,n)/l$, we have now
${\Gtilde}_\parallel = l \G_{\parallel}$.
  As a consequence, to a given value $k_\parallel$ of the quantum
number of the ribbon correspond $l$ values $(\tilde
k^{(0)}_\parallel,\cdots,\tilde k^{(l-1)}_\parallel)$ in the Brillouin
zone ($\tilde k^{(j)}_\parallel= k_\parallel + j
|{\Gtilde}_{\parallel}|)$, and therefore $l$ Zak phases.  The
prescription we shall use for non-coprime integers $(m,n)$ is
therefore that for a given value $k_\parallel$ of the ribbon quantum
number, the Zak phase is defined as
\begin{equation} \label{eq:noncoprime}
\zak_{(m,n)}(k_\parallel) \equiv \sum_{j=1}^l \zak_{(\tilde m,\tilde
  n)}(\tilde k^{(j)}_\parallel) \; .
\end{equation}
This prescription merely corresponds to a folding of the Brillouin zone.

\subsection{Zigzag boundary conditions}
\label{sec:zzbc}

\begin{figure}[ht]
  \centering
  \includegraphics[width=9cm]{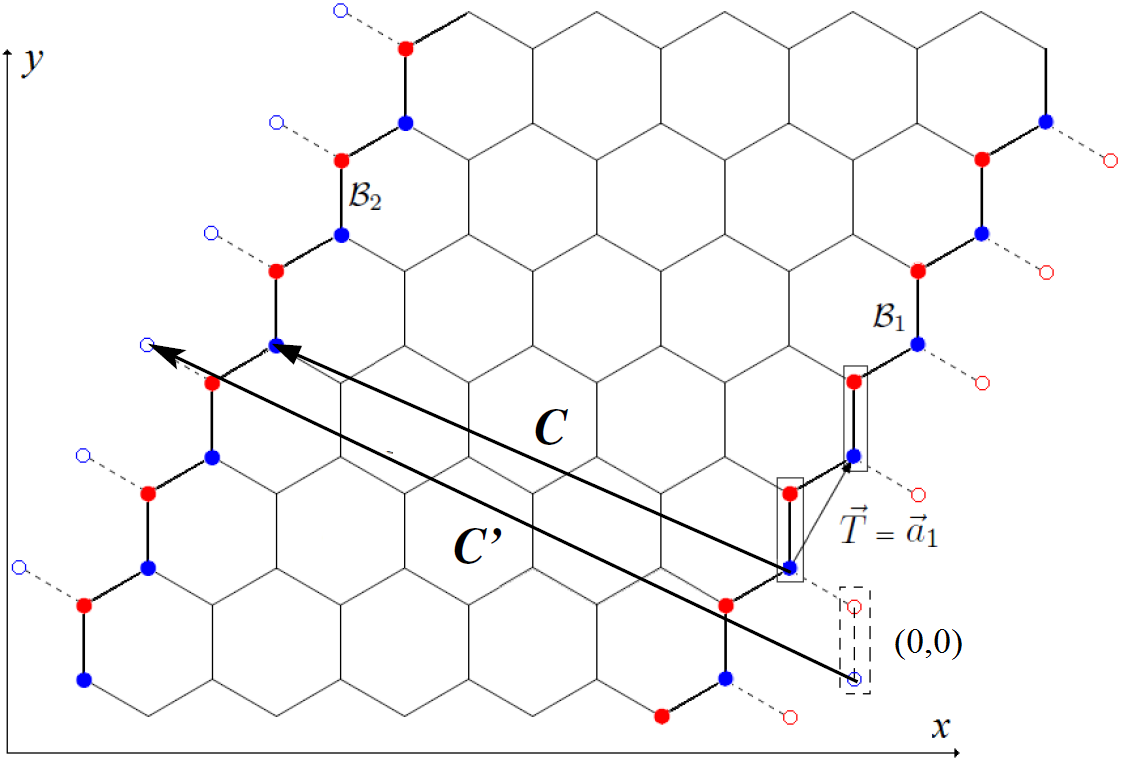}
  \caption{Schematic representation of a zigzag ribbon. The zigzag
    edge is obtained by translating the dimer $A$-$B$ (represented by
    a rectangle) with the periodicity $\T=\a_1$. The two edges
    $\mathcal{B}_1$ and $\mathcal{B}_2$ are represented by thick
    lines. The first vacant sites outside of the ribbon where we
    impose the wave function to vanish are represented by
    circles. $\C$ is the vector of the Bravais lattice which connects
    two sites on both edges, and $\C'$ connects the first vacant
      site on one side of the ribbon to the first vacant site on the
      other side.}
\label{fig:zig}
\end{figure}

Before we address general orientations for the graphene ribbon, let us
first consider in details the simple case of zigzag boundary
conditions.  In that case, and as illustrated in Fig.~\ref{fig:zig},
the two edges of the ribbon, $(\mathcal{B}_1)$ and $(\mathcal{B}_2)$, are constructed as the
vertical dimer $A$-$B$  translated periodically with the vector
 $\T = \a_1$ (see Fig.~\ref{fig:gph}). We note $\C$ the
vector which connects the two edges and $\C'= \C + 2\a_2 - \a_1$
the vector connecting $\mathcal{B}'_1$ and $\mathcal{B}'_2$, the lines
of empty sites nearest neighbors to the edges $\mathcal{B}_1$ and
$\mathcal{B}_2$.
Noting $\mathbf{\Psi}_{\k\pm} = (\Psi^A_{\k\pm},\Psi^B_{\k\pm})^T$ the Bloch
state of momentum $\k$, we have
\begin{equation}
\mathbf{\Psi_{\k,\pm}}(\r) =  \langle {\r} | \mathbf{\Psi}_{\k\pm} \rangle \propto
  \frac{e^{i {\k} \cdot {\r}}}{\sqrt{2}}
  \left( \begin{array}{c} e^{-i\phi(\k)} \\ \pm 1 \end{array} \right) \; .
\end{equation}
Following the approach used for the one-dimensional chain of dimers of
section~\ref{sec:open}, we would like to construct eigenstates of
the ribbon as a linear combination $\mathbf{\Psi}$ of two Bloch states
$\mathbf{\Psi}_{\k\pm}$ and $\mathbf{\Psi}_{\k'\pm}$ at the same energy.
From Fig.~\ref{fig:zig}, we see that the boundary conditions read
\begin{eqnarray}
\Psi^A(\C' + \nu {\T}) & = & 0 \qquad \mbox{[on   $\mathcal{B}'_2$]} \; ,
\label{eq:psi_B2}\\
\Psi^B(\nu {\T}) & = & 0 \qquad \mbox{[on $\mathcal{B}'_1$]} \; ,
 \label{eq:psi_B1}
\end{eqnarray}
with $\nu \in \mathbb{Z}$.

The boundary condition Eq.~(\ref{eq:psi_B1}) implies that the wave
functions $\mathbf{\Psi}$ are combinations of the form $\mathbf{\Psi} =\mathbf{\Psi}_{\k}
-\mathbf{\Psi}_{\k'}$ with ${\k}\cdot {\T} = {\k'}\cdot \T$ (here
${\T}={\a_1}$). Since we need $\epsilon_{\k} =
\epsilon_{\k'}$, we have to consider momentum pairs $({\k},{\k'})$ such
that $({\k}+{\k'})\cdot {\a_2} = {\k} \cdot{\a_1} = {\k'}
\cdot{\a_1}$.  For a given value of $k_\parallel$, it can be checked
that this is satisfied if
\begin{eqnarray}
{\k}   & = & {\k_\parallel} + {\k_\perp} \; ,\\
{\k'}  & = & {\k_\parallel} - {\k_\perp} \; ,
\end{eqnarray}
where we have introduced $ \k_\parallel = k_\parallel \e_\parallel$
and $ \k_\perp = k_\perp \e_\perp$.  Note that ${\k_\perp}=0$ or
${\k_\perp}={\G_\perp}/2$ correspond to $\mathbf{\Psi}_{\k} -\mathbf{\Psi}_{\k'} \equiv
0$.

The boundary condition Eq.~(\ref{eq:psi_B1}) then imposes
\begin{equation}
({\k} \! - \! {\k'}) \cdot \C' - (\phi({\k}) -
\phi({\k'})) = 2 \kappa \pi
\end{equation}
(for integer $\kappa$),
or in term of the phase $\phi(k_\parallel, k_\perp)= \frac{1}{2}
[\phi(\k_\parallel + \k_\perp) - \phi(\k_\parallel - \k_\perp)]$
\begin{equation}
\k_\perp \cdot \C' - \phi(k_\parallel,k_\perp)= \kappa \pi \; ,
\end{equation}
which is the strict equivalent of the quantization condition
(\ref{eq:qc}) obtained for the 1D chain. We note furthermore that
  $ (\G_\perp \cdot \C') = 2\pi(M+1)$, with $M$ the number of dimers
  in the transverse direction of the ribbon, and thus half the number
  of bands. Using  the same arguments as in section
  \ref{sec:open}, we see that the number of  edge states depends on the value of
  $\phi(k_\parallel,\Gamma_\perp/2)$, and more precisely  that the quantity
\begin{equation}
\left|\frac{\zak(k_\parallel)}{\pi}\right| =
  \left|\frac{1}{2\pi}
    \oint dk_\perp {\frac{d
        \phi(k_\parallel,k_\perp)}{dk_\perp}}\right|
\end{equation}
gives the number of pairs of edge states (with opposite energies) for a given $k_\parallel$.

\subsection{General orientation}
\label{sec:general}

We consider now graphene ribbons characterized by an arbitrary
translation vector ${\T}(m,n)$. Based on similar arguments as in the
previous section we conjecture that the relation between the Zak phase
and the number of edge states holds in this general case. Then we use
this relation to predict the existence of edge states in graphene
ribbons of general orientation.  We also check that for every
case for which edge states has been computed (numerically or
otherwise) their appearance is correctly predicted by the Zak phase.

\begin{figure}[ht]
  \centering
  \includegraphics[width=9cm]{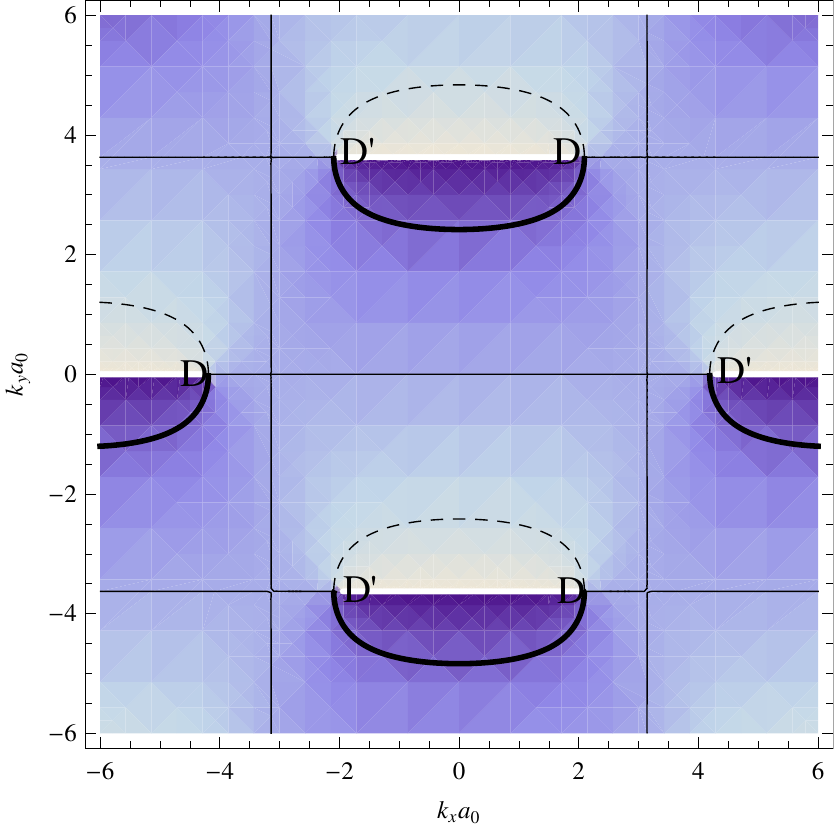}
  \caption{Density plot of the phase $\tilde{\phi}(\k)$. The discontinuities of the phase are shown by
    horizontal white lines connecting pairs of Dirac points. These
    singularities separate the values $\tilde{\phi}=+\pi$ (light
    region) to the value $\tilde{\phi}=-\pi$ (dark
    region). Consequently, the only paths $\Gamma_\perp$ that
    contribute to a non-vanishing Zak phase are those that cross these
    singularities. The black lines, thick lines and dashed lines
    represent the iso-$\tilde{\phi}$ lines respectively for
    $\tilde{\phi}=0$, $\tilde{\phi}=-\pi/2$ and
    $+\pi/2$. }
	\label{fig:densityplot}
\end{figure}

In Eq.~(\ref{eq:zakII}), the phase $\phi(\k)$ should be understood as
a multivalued function.  The single-valued function $\tilde \phi(\k)$ corresponding to
the restriction of $\phi(\k)$ to the interval $[-\pi,+\pi]$ is
displayed on Fig. \ref{fig:densityplot}. It shows lines of
discontinuity connecting pairs of Dirac points.
The location of the discontinuities (i.e.~which Dirac points are connected by them)
depends on the choice  made for the
unit cell dimer $A$-$B$ (oriented along the $y$ axis in this paper, see Fig.~\ref{fig:gph}).
Therefore the phase $\phi(\k)$ is not invariant by a rotation of
an angle $\pm2\pi/3$.

Let us start again by considering ribbons of period ${\T}(m,n)$ with
coprime $(m,n)$. The discontinuities of $\tilde \phi(\k)$ are
extremely convenient to determine the Zak phase since paths that lead
to a non-vanishing Zak phase necessarily  cross a discontinuity line.
Actually, the Zak phase $\zak_{(m,n)}(k_\parallel)$ is given by the
number of discontinuities $d(k_\parallel)$ intersected by the path
$\mathcal{P}_{(m,n)}(k_\parallel) = [\k_\parallel, \k_\parallel +
\G_\perp]$ along which $\zak_{(m,n)}(k_\parallel)$ is computed, that
is
\begin{equation} \label{eq:dkpar}
\zak(k_\parallel)=\pm \pi \ d(k_\parallel) \ .
\end{equation}
In other words, $d(k_\parallel)$  is just the number of pairs of edge states
(of opposite energies) for a given $k_\parallel$.

\begin{figure}[ht]
  \centering
  \includegraphics[width=9cm]{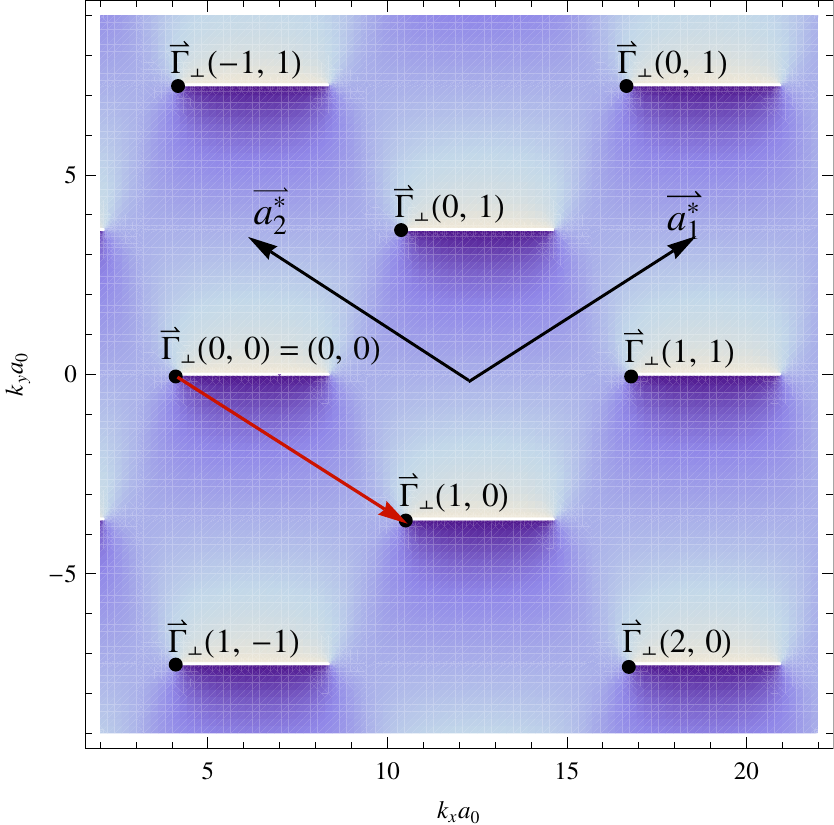}
  \caption{On top of the density plot of the phase  $\tilde{\phi}(\k)$, we have represented
  several values   of the  vector $\G_\perp(m,n)=n {\a}_1^\ast- m {\a}_2^\ast$.
  The left extremity of a discontinuity of $\tilde{\phi}(\k)$ is taken as the origin. We have
  explicitly plotted the vector  $\G_\perp(1,0)$ as an example.
   }
	\label{fig:map1}
\end{figure}

\begin{figure}[ht]
  \centering
  \includegraphics[width=9cm]{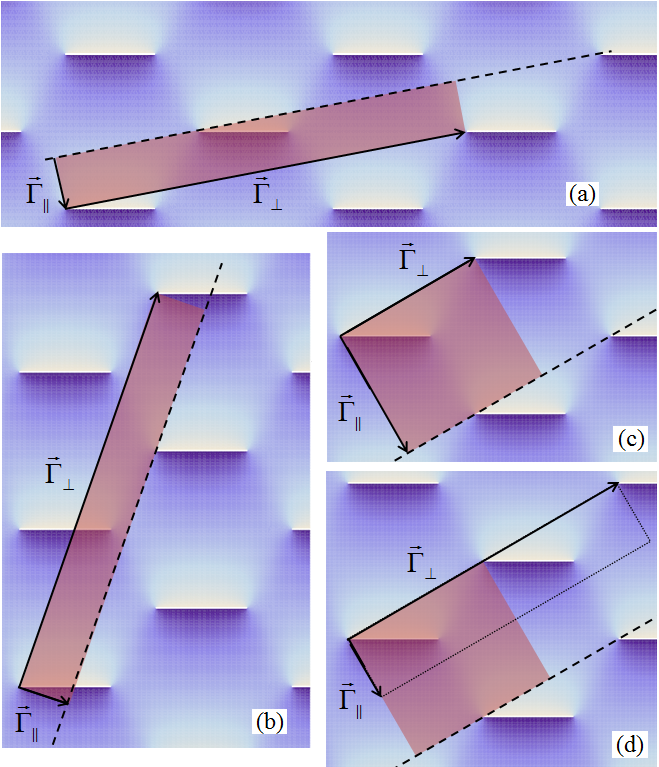}
  \caption{Surfaces $|{\G_\parallel}\wedge{\G_\perp}|$ associated with different
  ribbon vectors ${\T}(m,n)$.  $\G_\perp$ is obtained as
 $\G_\perp(m,n)=n {\a}_1^\ast- m {\a}_2^\ast=(n,-m)$; see text for the construction of $\G_\parallel$.
  (a) $\G_\perp(m=1,n=2)=(2,-1)$, (b) $\G_\perp(-2,3)=(3,2)$,
  and (c) $\G_\perp(0,1)=(1,0)$. In these three cases, $m$ and $n$
    are coprime, and the surfaces $|{\G_\parallel}\wedge{\G_\perp}|$ represented by a shaded rectangles
     are Brillouin zones. (d) ${\T}(0,2)$~: the
    surface obtained by ${\G_\parallel}$ and ${\G_\perp}$ is not a
    Brillouin zone because $m$ and $n$ are not coprime. The
    corresponding Brillouin zone (shaded rectangle) is given by
    ${\G_\perp}(0,1)={\G_\perp}(0,2)/2$ and ${\G_\parallel}(1,0)=2
    {\G_\parallel}(2,0)$.}
  \label{fig:bz}
\end{figure}

These considerations make it possible to compute graphically the Zak phase
in a rather straightforward way.
For a given choice of edge characterized by the vector ${\T}(m,n)= m {\a}_1 + n {\a}_2$, we first  represent the vector $\G_\perp(m,n)=n {\a}_1^\ast- m {\a}_2^\ast$ (Eq. \ref{gammaperp}). This is done in Fig.~\ref{fig:map1}, where we plot the vectors ${\G_\perp}(m,n)$
from the left extremity of the discontinuity $(0,0)$ to the left
extremity of the discontinuity $(n,-m)$. Next, in Fig.~\ref{fig:bz},  we translate
perpendicularly ${\G_\perp}$ until  the left extremity of an another
discontinuity is reached (dashed line), which gives
 ${\G_\parallel}(m,n)$.   The rectangle defined by
${\G_\perp}$ and ${\G_\parallel}$ is the Brillouin zone we want to associate with the ribbon.
For a given value of the momentum $k_\parallel$, $\zak(k_\parallel)$ is then deduced from Eq.~(\ref{eq:dkpar})
by simply counting the number of intersections of the segment $[\k_\parallel, \k_\parallel +
\G_\perp]$ with the discontinuity lines of $\tilde \phi(\k)$.

If $m$ and $n$ are not coprime, as for instance in
Fig. \ref{fig:bz}(d), ${\G_\parallel}(m,n)$ and ${\G_\perp}(m,n)$
cannot be obtained in this way as they do not define a Brillouin zone.
Writing however $(m,n) = l (\tilde m, \tilde n)$ with $(\tilde m,
\tilde n)$ coprime, the Brillouin zone corresponding to $(\tilde m,
\tilde n)$ can be constructed as above, and one  has simply
${\G_\parallel}(m,n) = {\Gtilde_\parallel}(\tilde m,\tilde n)/l $ and
${\G_\perp}(m,n) = l {\Gtilde_\perp}(\tilde m,\tilde n)$, which basically amounts to
a folding of ${\Gtilde_\parallel}$ by a factor $l$. For a given value of $k_\parallel$, the Zak phase in then
given by Eq.~(\ref{eq:noncoprime}).

\subsection{Range of existence and density of edge states}

In this section we derive  the range
  $\Delta k_\parallel$ for which edge states exist, as well as the
  density of edge states for arbitrary boundary conditions thanks to
  the bulk-edge correspondence in terms of the Zak phase.

\begin{figure}[ht!]
\centering
\includegraphics[width=9cm]{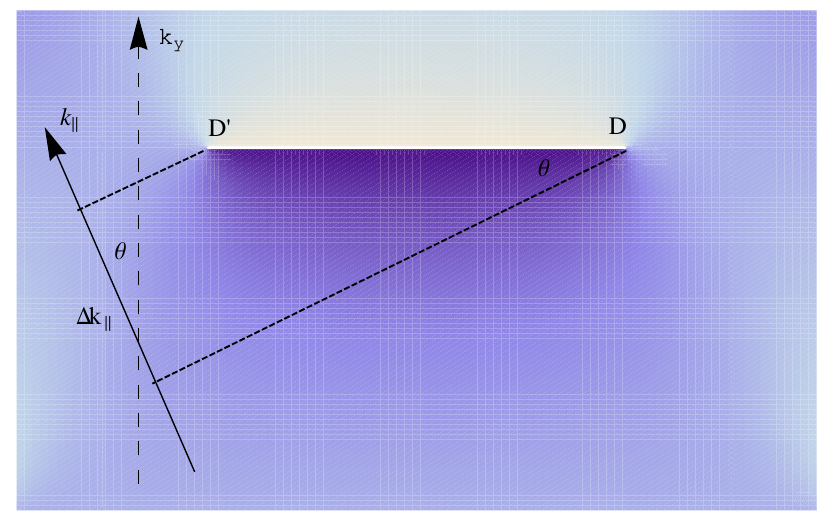}
\caption{Projection of the discontinuity of $\tilde{\phi}(\k)$
  onto the $k_\parallel$ axis. This projection gives the range $\Delta
  k_\parallel$ for the existence of edge states.}
	\label{fig:projection}
\end{figure}

In the last section, we showed that  the number of pairs of
edge states for a given value of
$k_\parallel$ is given by the number $d(k_\parallel)$ of crossings between the path
$\mathcal{P}_{(m,n)}(k_\parallel)$ and the discontinuities of
$\tilde{\phi}(\k)$. Since a Brillouin zone always contains
exactly one line of discontinuities,  the total range
\begin{equation}
\Delta k_\parallel \equiv \int_0^{2\pi/|\T|} d(k_\parallel)  d k_\parallel
\end{equation}
over which the ribbon exhibits edge states, is obtained by projecting
the line of discontinuities onto the $k_\parallel$ axis, as
illustrated in Fig. \ref{fig:projection}.  This leads to
\begin{equation} \Delta k_\parallel =
\frac{4\pi}{3a_0} \left| \sin \theta\right| \; ,
 \end{equation}
 where $\theta$ is the angle between the direction ${\T}$ of the
 ribbon and the vertical axis $y$ of the dimers (see Fig.~\ref{fig:gph} and Eq.~(\ref{theta})).  Therefore, there is no edge
 state for edges parallel to the armchair edge ($\theta=0$) and
 $\Delta k_\parallel$ is maximum and equal to $4\pi/3a_0$ for
 bearded edges.\cite{klein99,liu09}
Then, from Eq.~(\ref{theta}) we have  the relation between the
range $\Delta k_\parallel$ of   existence of edge states in graphene
and the integers $(m,n)$ characterizing the edge:
\begin{equation}
\Delta k_\parallel(m,n) =
\frac{2\pi}{3a_0}\frac{\left|n-m\right|}{\sqrt{n^2+m^2+nm}} \ .
\label{rangees}
\end{equation}
We notice that $\Delta k_\parallel(m,n)=\Delta k_\parallel(lm,ln)$
which means that ${\T}(m,n)$ and $l{\T}(m,n)$ support the same number
of localized states. A relevant quantity to study is the ratio ${\cal
  R}\equiv \Delta k_\parallel/|\G_\parallel|  =
 \Delta k_\parallel|\T|/ ( 2\pi)$   which gives the  relative range  of
the 1D Brillouin zone where edge states exist. We find
\begin{equation}
{\cal R}=\frac{\left|n-m\right|}{3} \ .
\label{r}
\end{equation}

We now comment the formulas (\ref{rangees}) and (\ref{r}).
First, we apply these formulas to several types of edges listed in table \ref{table1}.
\begin{table}[ht!]
\begin{tabular}{|c|c|c|c|}
\hline
               & $(m,n)$     &      $\Delta k_\parallel$  &   ${\cal R}$  \\
               \hline
\text{armchair}& $(1,1)$  & $0$ &  $0$\\
\hline
\text{zigzag}&  $(1,0)$  & $2\pi/3a_0$ & $1/3$\\
\hline
\text{bearded zigzag}& $(1,-1)$  & $4\pi/3a_0$ & $2/3$\\
\hline
\text{bearded armchair}& $(2,-1)$ & $2\pi/\sqrt{3}a_0$ & $1$\\
\hline
                      & $(1,3)$  &$4\pi/3\sqrt{13}a_0$& $2/3$\\
\text{general types}  & $(1,5)$ & $2\pi/3\sqrt{31}a_0$ & $4/3$ \\
                      & $(1,8)$   &  $14\pi/15\sqrt{3}a_0$ & $7/3$\\
                      \hline
\end{tabular}
\caption{Applications of the formulas (\ref{rangees}) and (\ref{r}) for some edge geometries studied in previous works.}
\label{table1}
\end{table}
The three general types of edges mentioned in this table represent a zigzag
profile extending over respectively three, five and eight unit cells followed by an
armchair defect (see for example Fig. \ref{fig:bords} for the case
$(1,5)$). These results are in good agreement with the size of the edge states energy bands obtained
by numerical tight-binding calculations in different previous works.\cite{fujita96,nakada96,hatsugai02,wakabayashi09,wakabayashi_jap10,jaskolski11}
We also notice that in the limit $|n|\gg |m|$ (or the other way), we
recover $\Delta k_\parallel\rightarrow 2\pi/3a_0$ which is the
expected result for the zigzag edge.

The formulas (\ref{rangees}) and (\ref{r}) directly lead to the
important result that edge states exist for most types of ribbons with a periodic pattern,
 which is in agreement with a previous analytical approach within the Dirac
framework.\cite{akhmerov} More precisely, we find that there is no
edge state if and only if $n=m$. As already mentioned,
this class of ribbons includes all the vectors ${\T}$ parallel to the
$y$ axis of the dimers $A-B$ (see for instance the case $(3,3)$ displayed in
Fig. \ref{fig:bords}. Similar ribbons have been synthesized for the first time recently\cite{mullen}).
 The well-known particular armchair case
corresponds to the smallest $|{\T}|$ that obeys this
condition.

\begin{figure}[ht]
  \centering
  \includegraphics[width=9cm]{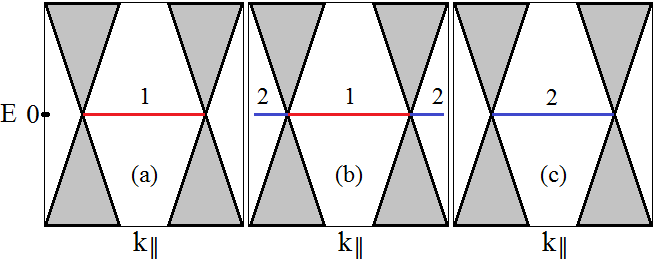}
  \caption{Schematic band structures of graphene ribbons that exhibit
    edge states. Edge states are represented as
      zero-energy flat bands as expected in the large width limit
    within the tight-binding
    model (assuming  the chiral symmetry
      is preserved). The   number of pairs of edge-states
        is indicated on the figure. (a) corresponds to
    the case  ${\cal R}<1$, (b) corresponds to ${\cal R} >1$. (c) Our
      analysis based on the Zak phase predicts that this
    configuration is not allowed.}
	\label{fig:deg}
\end{figure}

The case ${\cal R} \geq 1$ also deserves some attention. It
implies that two or more  pairs of localized states may correspond to the same
$k_\parallel$. This situation happens when, as illustrated in the
example of Fig. \ref{fig:bz}(b), several discontinuities are
intersected by the path $\mathcal{P}_{(m,n)}(k_\parallel)$. Such a situation is
automatically achieved for $m$ and $n$ coprime when
$\theta(m,n)>\theta(4,1)=\arctan(\sqrt{3}/5)$. Moreover, by
construction, the projection
  of the line of discontinuity of $\tilde \phi$ spans ${\cal R}$ times the
Brillouin zone, which  implies that the number of localized states cannot differ
by more than one unit for any two $k_\parallel$ (see Fig.~\ref{fig:deg}).
Therefore,  for $m$ and $n$ coprime, there is either
$[{\cal R}]$ or $[{\cal R}]+1$ edge states for each $k_\parallel$
($[x]$ is the floor function). In the particular case where ${\cal
  R}$ is an integer there are exactly ${\cal R}$ edge states for
each $k_\parallel$.

  Finally, we can define the quantity $\rho={\cal R}/|\T|$ which corresponds to
  the ``density of edge states per unit length''   introduced by Akhmerov and
  Beenakker.   We get
\begin{equation}
\rho=\frac{\Delta k_\parallel}{2\pi} =
\frac{1}{3a_0}\frac{\left|n-m\right|}{\sqrt{n^2+m^2+nm}} \; ,
\end{equation}
 which was first obtained  in Ref.~[\onlinecite{akhmerov}] for minimal
 boundary conditions by a different method.

\section{Emergence and destruction of edge states: a topological approach}
\label{sec:asym}

In this section, we generalize the formula (\ref{rangees}) for non-equal
hopping parameters $t_1 \neq t_2 \neq t_3$, and establish a
criterion for the existence of edge states for an anisotropic
honeycomb lattice.  We show that the manipulation of these parameters
leads to a topological transition described in terms of Zak phase that affects
the range $\Delta k_\parallel$ of existence of edge states.  We
  stress that breaking the isotropy of the hopping parameters
  preserves the chiral symmetry, and therefore the topological
  character of the Zak phase.  As a consequence the analysis
  developed in the previous section generalizes straightforwardly to
  the case considered below.

\subsection{Effect of an anisotropy on the existence of edge states}

\begin{figure}[ht!]
\centering
\includegraphics[width=8cm]{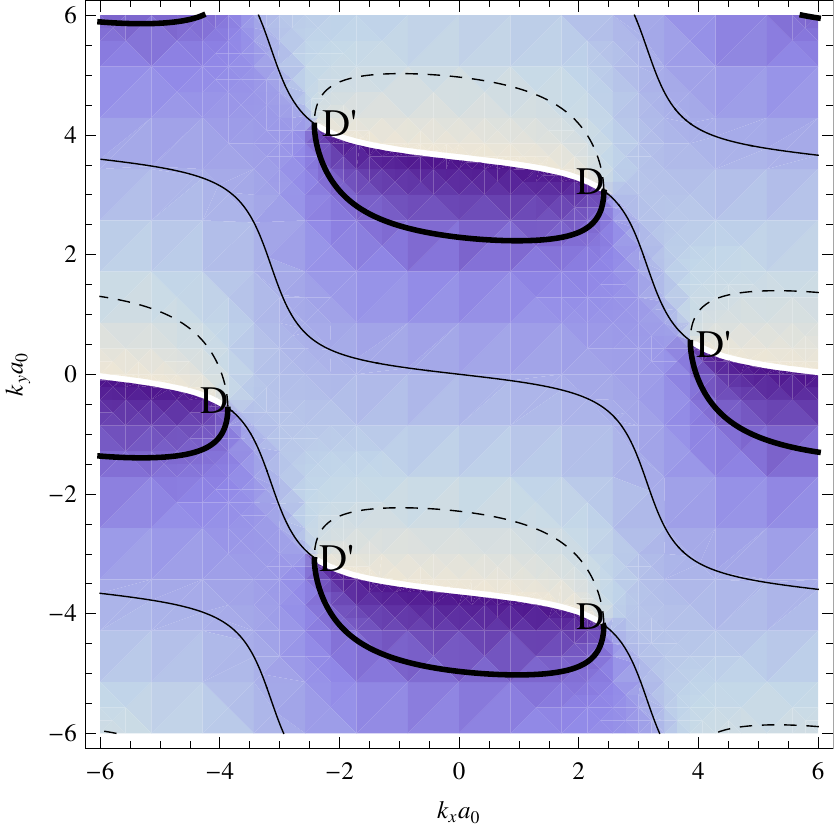}
\caption{Density plot of the phase $\tilde{\phi}({\k})$ for $t_2=1.5t_1=1.5t_3$. The discontinuities
    of the phase are shown by curved white lines that pair the
    Dirac points. The black lines, thick lines and dashed lines
    represent the iso-$\tilde{\phi}$ lines respectively for $\tilde{\phi}=0$,
    $\tilde{\phi}=-\pi/2$ and $+\pi/2$. }
\label{fig:asym1}
\end{figure}

\begin{figure}[htb]
\centering
\includegraphics[width=8cm]{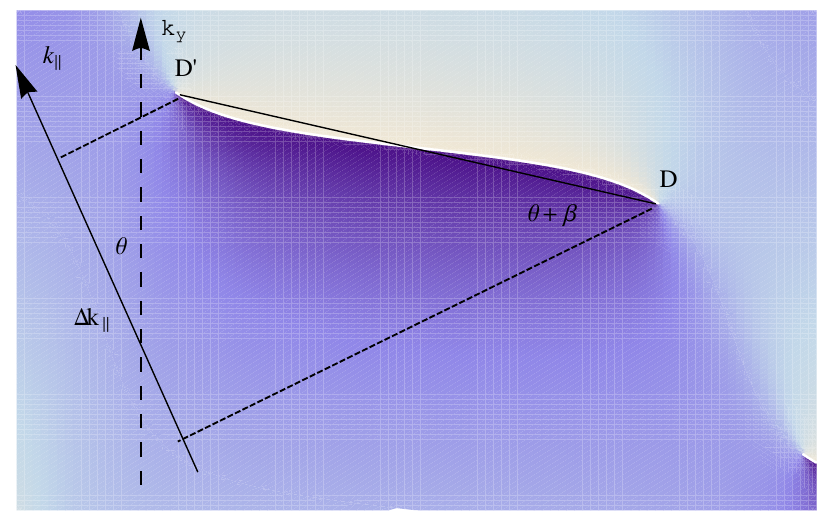}
\caption{Projection of the discontinuity line  of $\tilde \phi (\k)$
onto the $k_\parallel$ axis. The asymmetry of the hopping induces a
modification of the discontinuity locations  as compared
 with  Fig.~\ref{fig:projection}.}
\label{fig:projection2}
\end{figure}

Several previous works dealing with the tight-binding model in the
honeycomb lattice showed that the Dirac points move when modifying the
ratio of the parameters
$t_i/t_j$.\cite{hasegawa06,dietl08,guinea08,merginguniv09} As the lines of
discontinuities of $\tilde{\phi}(\k)$ connect pairs of Dirac points,
the  modification of the ratio $t_i/t_j$ changes the Zak phase and
therefore leads to new ranges $\Delta k_\parallel$  of existence
of edge states. This is clearly shown in Fig.~\ref{fig:asym1}.
To determine these new ranges, we have first to specify the position
of the Dirac points for any $t_i$. Up to a vector of the reciprocal
lattice, they are given by
\begin{equation}
\begin{split}
{\D} &=-\frac{\pi-d_1}{2\pi}{\a_1^\ast}+\frac{\pi-d_2}{2\pi}{\a_2^\ast} \\
{\D}' &=\frac{\pi-d_1}{2\pi}{\a_1^\ast}-\frac{\pi-d_2}{2\pi}{\a_2^\ast}
\label{dprime}
\end{split}
\end{equation}
with
\begin{equation}
\begin{split}
d_1 = &\mathbbm{R}e
\left[\arccos\left(\frac{t_3^2+t_2^2-t_1^2}{2t_2t_3}\right) \right]
 \\ d_2 = & \mathbbm{R}e
\left[\arccos\left(\frac{t_3^2+t_1^2-t_2^2}{2t_1t_3}\right) \right] \; ,
\end{split}
\label{d1d2}
\end{equation}
where $\mathbbm{R}e(x)$ takes the real part of $x$. The cartesian coordinates od the Dirac points
${\D}^{(\prime)}=(D_x^{(\prime)},D_y^{(\prime)})$ are given
by
\begin{equation}
  \begin{split}
    {\D}
    & = \left( \left( d_1+d_2-2\pi \right) /a_0, \left( d_1-d_2 \right) /
      \sqrt{3}a_0 \right)
    \\
    {\D'} & = \left( \left( -d_1-d_2+2\pi \right )/a_0 , \left( d_2-d_1
      \right )/\sqrt{3}a_0 \right)
    \; .
  \end{split}
  \label{dprime2}
\end{equation}
Then, since the Dirac points ${\D}^{(\bf{\prime})}$ are not located anymore at
the corner ${\K}^{(\bf{\prime})}$ of the Brillouin zone, the range
$\Delta k_\parallel$ of existence of edge states is modified as:
\begin{equation}
  \Delta k_\parallel = |\D - \D'|\left|\sin\left(\theta+\beta\right)\right|
\end{equation}
 where $\beta$ is
 the angle between the line $[{\bf D'} {\bf D}]$  and the $k_x$ axis (see
 Fig.~\ref{fig:projection2}), which is then given by
\begin{equation}
  \begin{split}
    \cos \beta = & 2\frac{2\pi/a_0-D_x^{\prime}}{ |\D- \D'|} \; ,\\
    \sin \beta = & 2\frac{D_y^{\prime}}{ |\D- \D'|} \; .
    \label{beta}
  \end{split}
\end{equation}

 Using Eqs.~(\ref{theta}) and (\ref{beta}), the range for the edge states reads:
\begin{eqnarray}
\Delta k_\parallel
 &=& \frac{\left|\left(n-m\right)\left(\frac{2\pi}{a_0}-D'_x\right)+\sqrt{3}\left(n+m\right)D'_y\right|}{\sqrt{n^2+m^2+mn}} \ .
\end{eqnarray}
Next, by using the expression of the positions $D'_x$ and $D'_y$ given in (\ref{dprime2}) one finds:
\begin{equation}
\Delta k_\parallel = \frac{2\left|  nd_2-md_1 \right|}{a_0\sqrt{n^2+m^2+nm}}
\label{rangees2}
\end{equation}
as well as
\begin{equation}
{\cal R}=\frac{\left|  nd_2-md_1 \right|}{\pi} \; ,
\label{r2}
\end{equation}
where $d_1$ and $d_2$ are given by Eqs.~(\ref{d1d2}).  These results
give a criteria for the existence of edge states that links the
anisotropy of the hopping parameters encoded in $d_1$ and $d_2$ with
the nature of the edge characterized by $(m,n)$. In the isotropic
case, we have $d_1=d_2=\pi/3$, and we recover the result
(\ref{rangees}) discussed in the previous section. The formula
(\ref{rangees2}) means that an edge state exists in graphene-like
structures if $\left| nd_1-md_2 \right|\neq 0$.

\begin{figure}[hbt]
\centering
\includegraphics[width=4cm]{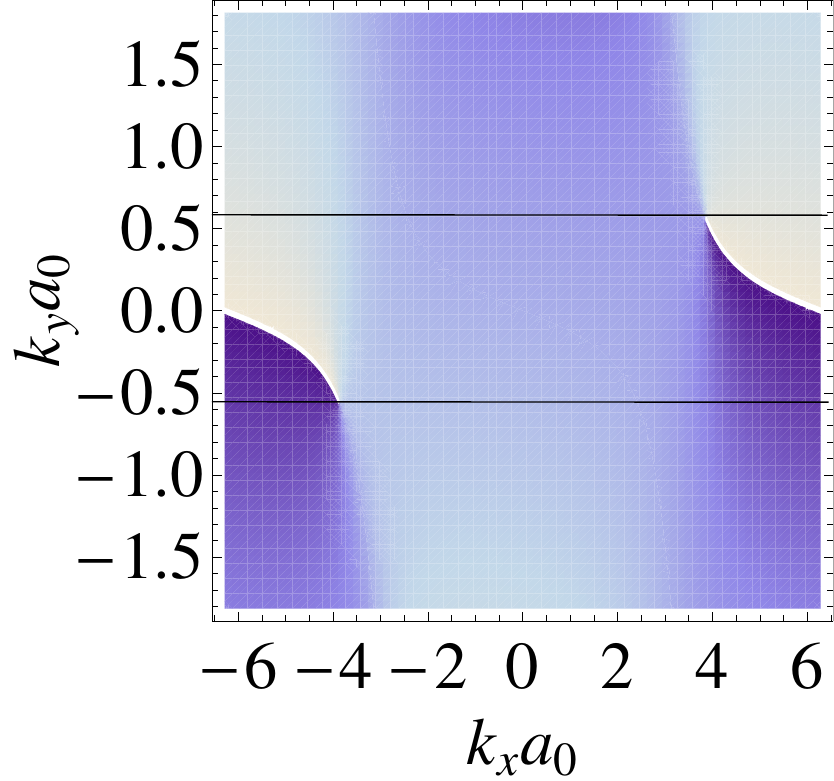}
\includegraphics[width=5cm]{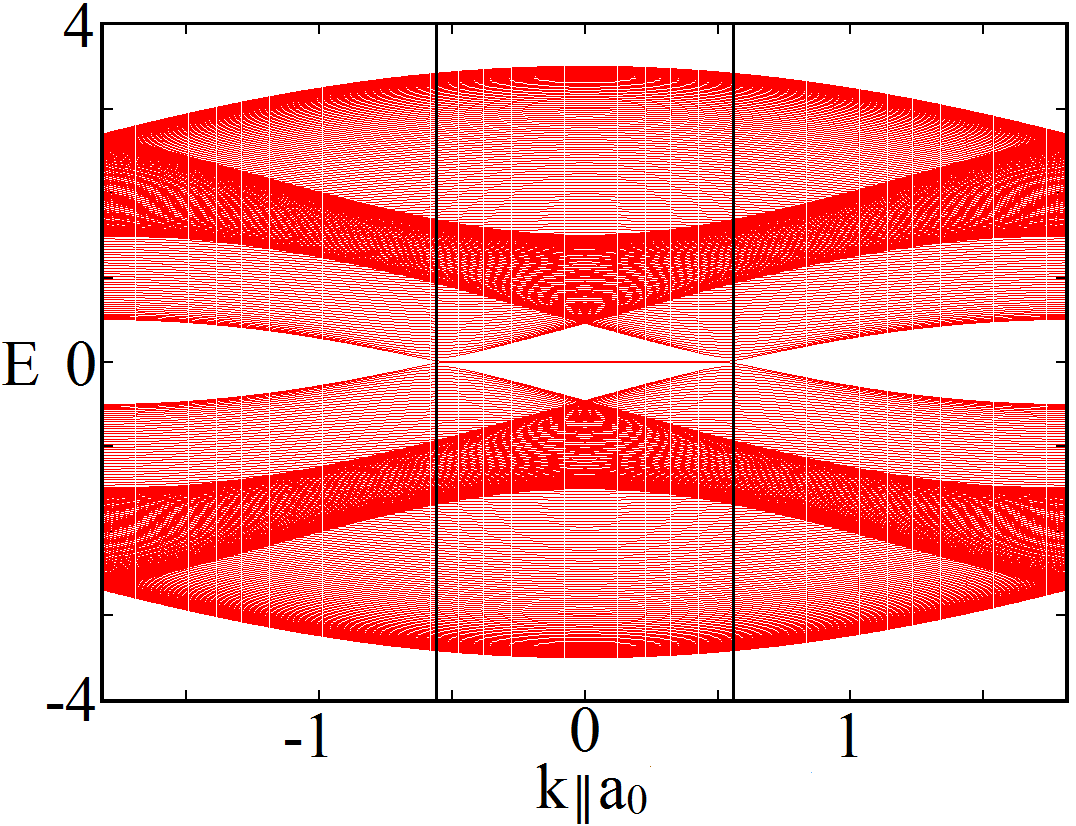}
\caption{(Left) Density plot of the phase ${\tilde \phi}(\k)$ for $t_1=1$, $t_2=1.5$ and
  $t_3=1$ represented in the Brillouin zone corresponding to armchair
  edges. The two horizontal lines delimit the region  where ${\cal Z}(k_\parallel)=\pi$.
  (Right) Band structure of an armchair ribbon with the same hopping
  parameters. In this case, $k_\parallel=k_y$. The two vertical lines delimit the same range
  as in the left panel, which is   such that edge states at zero energy,
  clearly separated from the bulk   bands, have emerged.  }
\label{fig:armmodif}
\end{figure}

 An interesting consequence is that, for a given
type of ribbon, edge states can emerge or collapse  when an anisotropy is applied. For instance, edge states
can emerge for armchair-like boundary conditions $(m=n)$ when either
$t_1/t_3\neq 1$ or $t_2/t_3\neq 1$ (see Refs.~[\onlinecite{dahal10}] and [\onlinecite{delplacethese}] for the case
$n=m=1$).  This is clearly displayed in
Fig. \ref{fig:armmodif}.
In the same way, edge states can collapse by manipulating
 the asymmetry of the hopping in such a way
  that $\left| nd_1-md_2 \right|=0$.

\subsection{Merging of Dirac points and edge states}

As one increases the anisotropy of the system, for instance by
modifying one of the ratio $t_i/t_j$, one may eventually reach a point
where $t_1 = t_2 + t_3$ (or the equivalent up to a cyclic
permutation of the indices).    From  Eq.~(\ref{d1d2}), this
condition implies that   $d_1$ and $d_2$
 take the values $0$ or $\pi$, which,  from
 Eq.~(\ref{dprime2}), corresponds to  {\it a merging of the Dirac points}.
 By increasing further the anisotropy, a gap opens at the merging point.
 This merging is a topological transition since
the Berry phases $\pm \pi$ associated to the two Dirac points   annihilate at the
 transition.\cite{dietl08,merginguniv09}

\begin{figure}[htb]
\centering
\includegraphics[width=4cm]{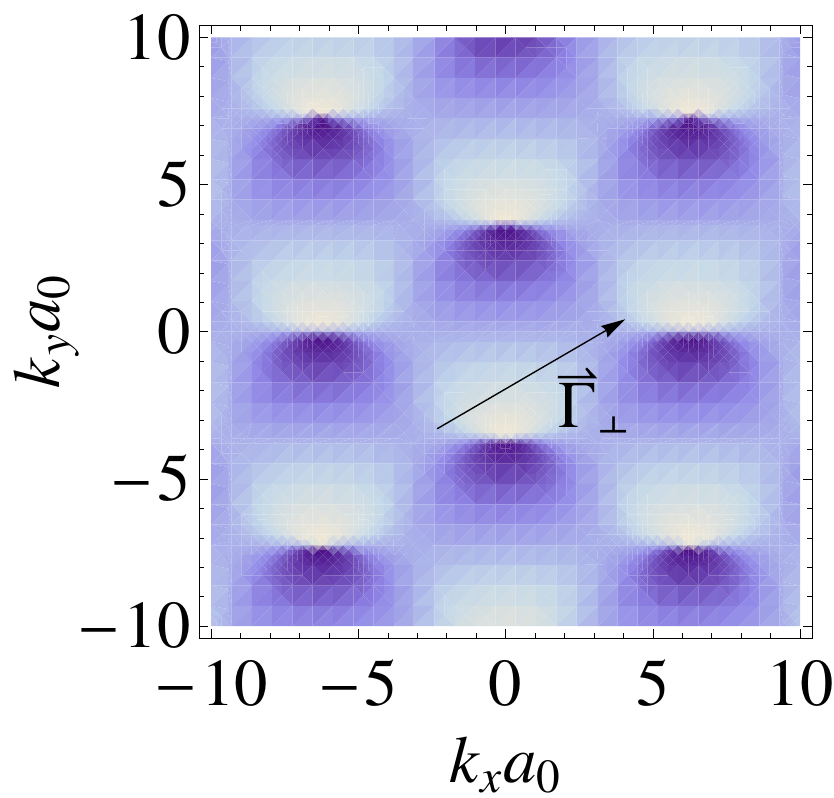}
\includegraphics[width=5cm]{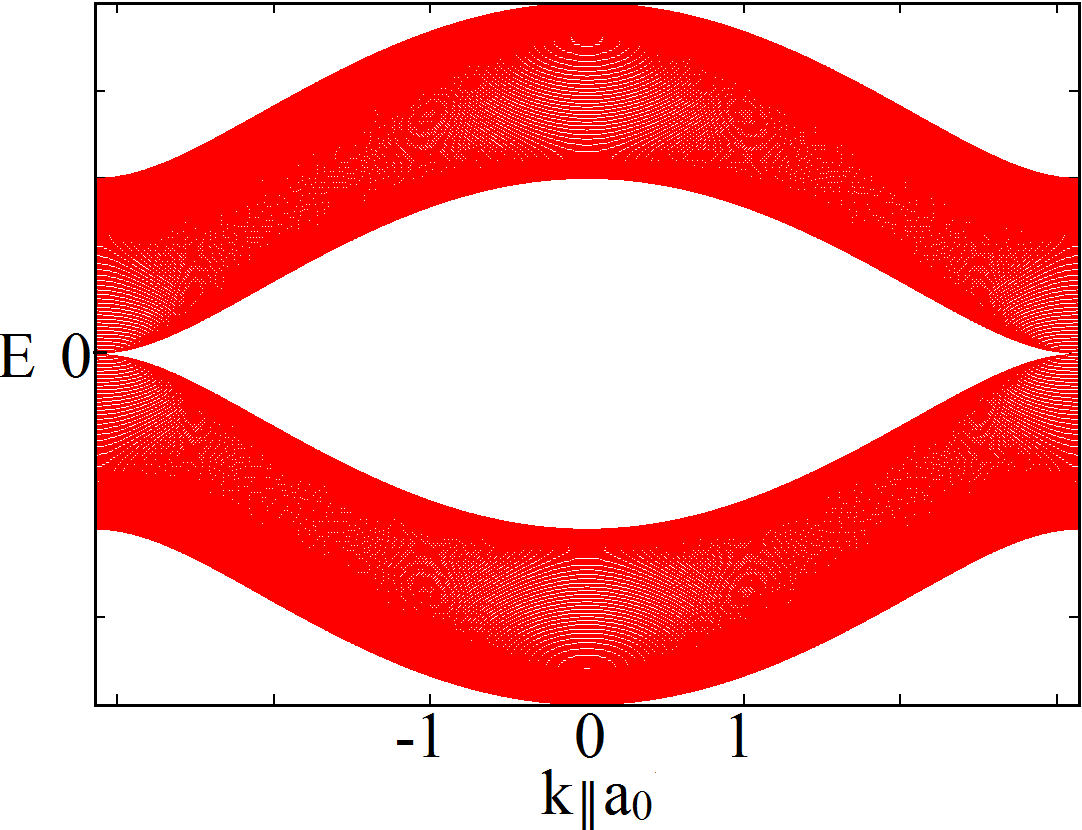}
\includegraphics[width=4cm]{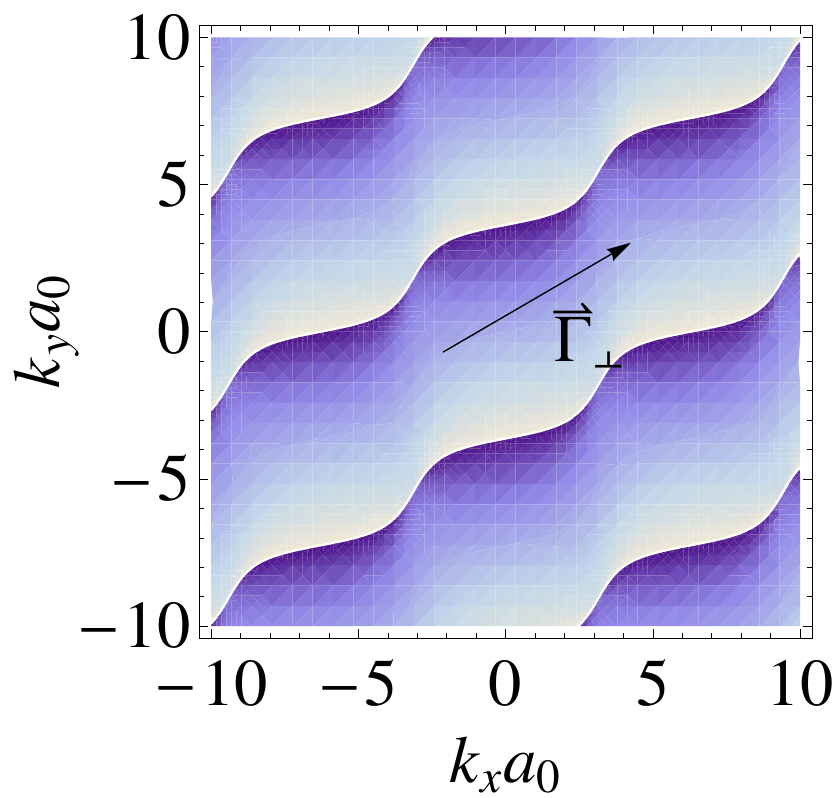}
\includegraphics[width=5cm]{zz_1_2_1}
\includegraphics[width=4cm]{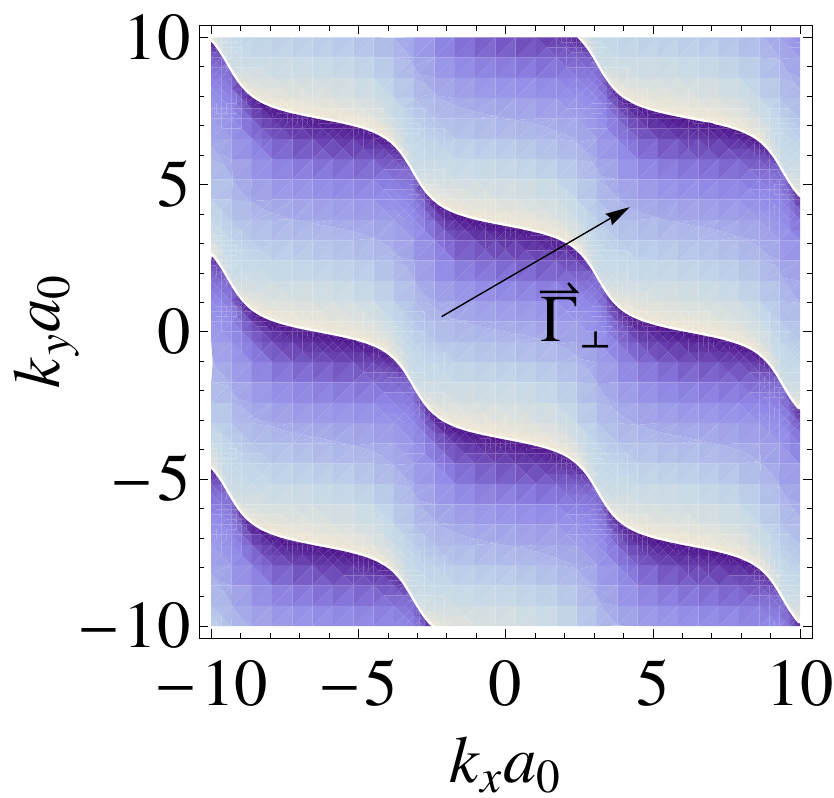}
\includegraphics[width=5cm]{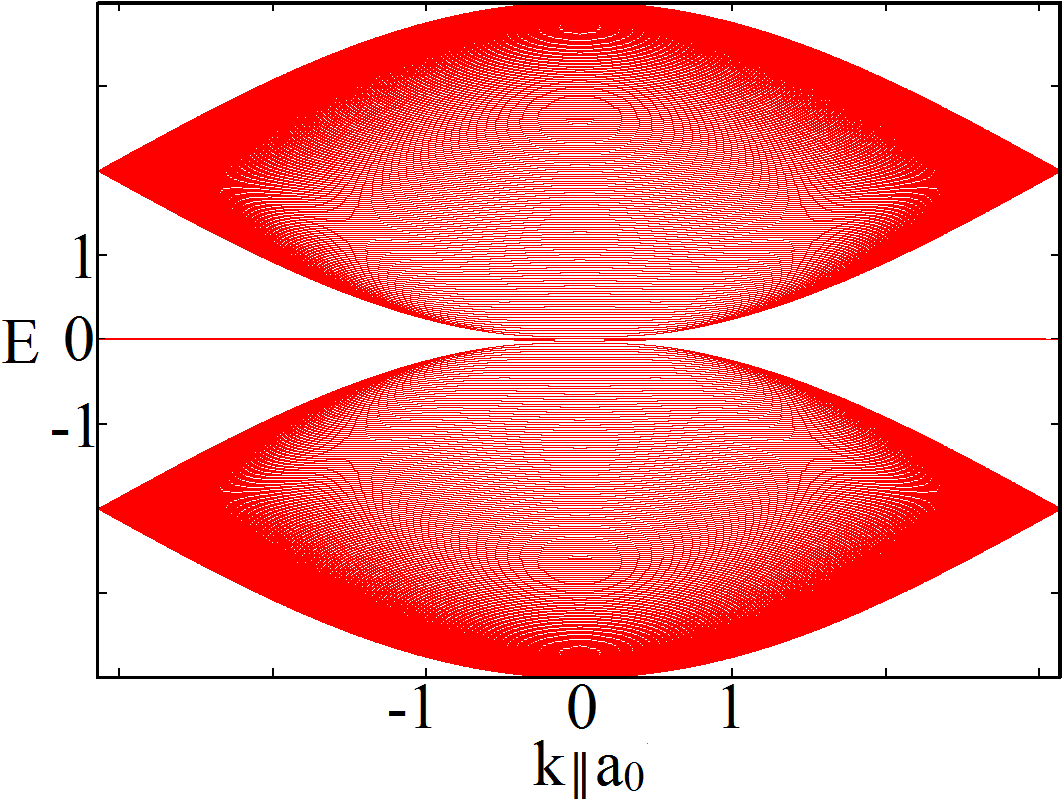}
\caption{Density plot of the phases ${\tilde \phi}(\k)$ at the merging
  of the Dirac
  points, and band structure for zigzag ribbon with $\theta=\pi/6$
  (that is ${\T}(m=0,n=1)\rightarrow
  {\G}_\perp(m=0,n=1)=(1,0)$). (Top) $t_1=1$, $t_2=1$ and $t_3=2$,
  (center) $t_1=2$, $t_2=1$ and $t_3=1$, (bottom) $t_1=1$, $t_2=2$ and
  $t_3=1$ . The scale is given in units of $1/a_0$. }
\label{fig:modif}
\end{figure}

This topological transition is a bulk property, and is thus independent
of the orientation of the anisotropy. 
 On the other hand, the existence of edge states at the merging transition and beyond
 depends on the orientation of the ribbon with respect to this  anisotropy, and this
 information is still contained in the Zak phase.  Indeed the anisotropy of the hopping
  parameters controls the size and the location of the lines of
  discontinuities of ${\tilde \phi}(\k)$. It is therefore essential to
  distinguish which pair of Dirac points merges. There are three
  possibilities :

\begin{itemize}

\item $t_3 \geq t_1+t_2$

  In this configuration, the Dirac points that merge are the endpoints
  of the same  discontinuity line. Then, the discontinuities
  disappear at the merging ($t_1+t_2=t_3$) and the Zak phase
  vanishes. This leads to
\begin{equation}
{\cal R}=0
\end{equation}
whatever $m$ and $n$ (see  top panel of Fig.~\ref{fig:modif}),
implying that  edge states never exist in this case. As an illustration, we plot in Fig.~\ref{fig:modif} the band
structure of a zigzag ribbon exactly at the merging point: the zero
energy edge-states\cite{fujita96} of the isotropic case have collapsed
because the Zak phase is zero for all $k_\parallel$.
\end{itemize}

The situation is totally different when the two merging Dirac points
are attached to two distinct discontinuity lines, which occurs when
$t_1-t_2=\pm t_3$. In this case, the two discontinuity-lines
themselves merge, implying that ${\cal R}$ is an integer, as we
discuss now.

\begin{itemize}
\item $t_1\geq t_2+ t_3$

In this case we have $d_2=0$ and $d_1=\pi$,  which leads to
\begin{equation}
{\cal R}=|m|
\end{equation}
for all $n$.  For instance, there is no edge state for
zigzag edge at $\theta=+\pi/6$ (see center panels of Fig.~\ref{fig:modif}
), but there is one  over  the whole ribbon
Brillouin zone  for armchair edges and for zigzag edges at
$\theta=-\pi/6$. We notice that the existence of edge states now only
depends on the orientation of the edge given by $\theta$ but not on
$k_\parallel$ anymore.

\item $t_2\geq t_1 + t_3$

In this case $d_1=0$ and $d_2=\pi$ and we find
\begin{equation}
{\cal R}=|n| \,
\end{equation}
	
whatever the value of $m$. Of course this case is equivalent to the
one discussed previously with the substitution $m\leftrightarrow n$
that is $\theta \rightarrow -\theta$. The phase $\tilde \phi(\k)$ and
the corresponding band structure for zigzag ribbons at $\theta=+\pi/6$
are represented in Fig. \ref{fig:modif} (bottom panels).
\end{itemize}

In summary, we note that in all three cases, at the merging,
  the Zak phase becomes independent of $k_\parallel$ and remains
  unchanged in the gapped phase. This implies that at the merging transition and beyond, ${\cal R}$ is
  necessarily an integer.    This is obvious from Figs.~\ref{fig:modif} : the merging of the Dirac points implies either a disparition  of the discontinuity lines or their transformation into an infinite line.
\section{Conclusion}

In this paper, we have investigated the correspondence between the Zak
phase and the existence of edge states for arbitrarily oriented graphene ribbons with a large class of edge shapes.
 We have proposed
  a definite prescription to compute the Zak phase in order to
  predict the number of edge states. The approach we have
  developed consists in constructing the appropriate 2D Brillouin zone
  associated with the vector ${\T}(m,n)$ which defines the edge. The
  Zak phase $\zak(k_\parallel)$ giving the number of edge states for
  each $k_\parallel$ is then directly obtained by integrating the
  Berry connection along a path fixed by
  $\T(m,n)$ and $k_\parallel$ in this 2D Brillouin zone.

  We stress that this bulk-edge correspondence is, beyond the 1D chain
  of dimers case, only rigorously proven here for zigzag edges.  It is
  therefore so far a conjecture for the class of edges we have
  defined. This conjecture is however supported by the fact that it
  reproduces all the known previous results obtained (numerically or otherwise) in the literature for various specific types of edges.
  \cite{fujita96,nakada96,hatsugai02,wakabayashi09,wakabayashi_jap10,jaskolski11}

In practice, the value of the integral defining the Zak phase is
easily obtained graphically.  Our approach therefore does not require
any sophisticated formalism or calculation, and gives an elegant
understanding of the origin of the edge-dependent edge states in terms
of a topological bulk quantity.  In particular, it provides a simple understanding  of
 the appearance and disappearance of edge
states by manipulating the anisotropy of the tight-binding hopping
parameters. Such a manipulation may be induced in graphene by applying
an uniaxial stress or bending of the sheet,\cite{pauling} or in photonic crystals
which mimic the same physics, by changing the distance between the confining mirrors.\cite{kuhl}

We finish with a few comments concerning the connection between
Zak phase and edge states.

First, this bulk-edge correspondence differ from  the ones in quantum Hall
systems\cite{hatsugai93,hatsugaiPRL93,hao,qi06} or $\mathbb{Z}_2$
topological insulators\cite{wang,qi06} since here the existence of
edge states precisely depends on the orientation
of the edge.  This
difference with the usual bulk topological numbers originates from the
fact that the Zak phase is a 1D  (rather
than 2D) integral of the Berry connection.

Second, we stress that within our approach, the vector ${\T}(m,n)$
defining the periodicity of the ribbon entirely determine the Zak
phase.  As many different shapes may correspond to the same vector
${\T}(m,n)$, the Zak phase and therefore the number of edge states are
expected to be independent of the variation of the edge geometries as
long as they correspond to the same ${\T}(m,n)$.\footnote{Preliminary numerical results seem to confirm
this prediction, (L. Bilteanu and C. Bena, Private communication).}

Finally, our description of edge states in terms of the Zak phase is a
priori not restricted to graphene but should be in principle also
applicable to other 2D systems like d-wave
superconductors,\cite{hatsugai02} square lattice with half a quantum
flux per unit cell\cite{delplacethese} or bi-layer
graphene\cite{jianli_prb10} for instance.

{\it Acknowledgments - } We acknowledge useful discussions with J.-N Fuchs, M. B\"uttiker, J. Li, C. Bena and L. Bilteanu.
This work is supported by the NANOSIMGRAPHENE Project No. ANR-09-NANO-016-01 of ANR/P3N2009.
In Geneva P. D. was supported by the European Marie Curie ITN  NanoCTM.

\appendix

\section{Construction of the Brillouin zone}
\label{app:BZ}

In this appendix, we give a brief reminder of the reason why the 
vectors $\G_\parallel$ and $\G_\perp$ defined by
Eqs.~(\ref{gammaparall})-(\ref{gammaperp}) actually generate a
Brillouin zone when $n$ and $m$ are coprime integers.

This latter condition indeed  implies that
one can find two integers $(m',n')$ such that $mn' - nm'=1$, in which
case the couple of vector $(\T,{\bf N})$, with ${\bf N} = m'
\a_1 + n' \a_2$, form a basis of the Bravais lattice.  The
choice of $(m',n')$, and thus of ${\bf N}$, is not unique, but this is
irrelevant for our purpose.

From $(\T,{\bf N})$, one deduce a  basis
$(\G_N,\G_\perp)$ 
of the reciprocal lattice,
\begin{eqnarray}
\G_N & = &  n'\a^*_1 - m'\a^*_2 \; , \\
\G_\perp    & = &  n \a^*_1 - m \a^*_2 \; ,
\end{eqnarray}
which is such that $\G_\perp \perp \T$.  A Brillouin
zone can thus be obtained from the parallelogram generated by
$(\G_N,\G_\perp)$.  More generally however, any
vector $\G$ such that
$(\G-\G_N) \parallel \G_\perp$ is such
that the parallelogram generated by $(\G
,\G_\perp)$ is a Brillouin zone.  A
natural choice is to take for $\G$ the vector
$\G_\parallel$ which is parallel to $\T$ (and thus
orthogonal to $\G_\perp$).  Since $\T \cdot
\G_N = 2\pi (mn'-nm') = 2\pi$, one has
$|\G_\parallel| = 2\pi/|\T| $ which is nothing but the
size of the (1D) Brillouin zone of the ribbon.

\bibliographystyle{phreport}
\bibliography{bibliob}

\begin{thebibliography}{10}

\bibitem{novoselov}
A.~Geim and K.~Novoselov,
\newblock The rise of graphene,
\newblock Nature materials {\bf 6}, 183 (2007).

\bibitem{fujita96}
M.~Fujita, K.~Wakabayashi, K.~Nakada, and K.~Kusakabe,
\newblock Pecular localized state at zigzag graphite edge,
\newblock J. Phys. Soc. Jpn {\bf 65}, 1920 (1996).

\bibitem{nakada96}
K.~Nakada, M.~Fujita, G.~Dresselhaus, and M.~S. Dresselhaus,
\newblock Edge state in graphene ribbons: Nanometer size effect and edge shape
  dependence,
\newblock Phys. Rev. B {\bf 54}, 17954 (1996).

\bibitem{kobayashi}
Y.~Kobayashi, K.~I. Fului, T.~Enoki, K.~Kusakabe, and Y.~Kaburagi,
\newblock Observation of zigzag and armchair edges of graphite using scanning
  tunneling microscopy and spectroscopy,
\newblock Phys. Rev. B {\bf 71}, 193406 (2005).

\bibitem{niimi}
Y.~Niimi, T.~Matsui, H.~Kambara, K.~Tagami, M.~Tsukada, and H.~Fukuyama,
\newblock Scanning tunneling microscopy and spectroscopy of the electronic
  local density of states of graphite surfaces near monoatomic step edges,
\newblock Phys. Rev. B {\bf 73}, 085421 (2006).

\bibitem{kuhl}
U.~Kuhl, S.~Barkhofen, T.~Tudorovskiy, H.-J. Stockmann, T.~Hossain,
  L.~de~Forges~de Parny, and F.~Mortessagne,
\newblock Dirac point and edge states in a microwave realisation of
  tight-binding graphene-like structures,
\newblock Phys. Rev. B {\bf 82}, 094308 (2010).

\bibitem{louie}
Y.-W. Son, M.~L. Cohen, and S.~G. Louie,
\newblock Half-metallic graphene nanoribbons,
\newblock Nature {\bf 444}, 347 (2006).

\bibitem{yazyev}
O.~V. Yazyev and M.~I. Katsnelson,
\newblock Magnetic correlations at graphene edges: Basis for novel spintronics
  devices,
\newblock Phys. Rev. Lett {\bf 100}, 047209 (2008).

\bibitem{halperin}
B.~I. Halperin,
\newblock Quantized Hall conductance, current-carrying edge states, and the
  existence of extended states in a two-dimensional disordered potential,
\newblock Phys. Rev. B {\bf 25}, 2189 (1982).

\bibitem{buttiker}
M.~B\"uttiker,
\newblock Absence of backscattering in the quantum Hall effect in multiprobe
  conductors,
\newblock Phys. Rev. B {\bf 38}, 9375 (1988).

\bibitem{bernevig06}
B.~A. Bernevig, T.~A. Hughes, and S.-C. Zhang,
\newblock Quantum spin Hall effect and topological phase transition in HgTe
  quantum wells,
\newblock Science {\bf 314}, 1757 (2006).

\bibitem{konig08}
M.~Konig, H.~Buhmann, L.~W. Molenkamp, T.~Hughes, C.-X. Liu, X.-L. Qi, and
  S.-C. Zhang,
\newblock The quantum spin Hall effect: theory and experiment,
\newblock J. Phys Soc. Jpn {\bf 77}, 031007 (2008).

\bibitem{TKNN}
D.~J. Thouless, M.~Kohmoto, M.~P. Nightingale, and M.~den Nij,
\newblock Quantized Hall conductance in a two-dimensional periodic potential,
\newblock Phys. Rev. Lett. {\bf 49}, 405 (1982).

\bibitem{hatsugai02}
S.~Ryu and Y.~Hatsugai,
\newblock Topological origin of zero-energy edge states in particule-hole
  symmetric systems,
\newblock Phys. Rev. Lett. {\bf 89}, 077002 (2002).

\bibitem{mong10}
R.~Mong and V.~Shivamoggi,
\newblock Edge states and the bulk-boundary correspondence in Dirac
  Hamiltonians,
\newblock Phys. Rev. B {\bf 83}, 125109 (2011).

\bibitem{sasakiNJP10}
K.~Sasaki, K.~Wakabayashi, and T.~Enoki,
\newblock Berry's Phase for Standing Wave Near Graphene Edge,
\newblock New J. Phys. {\bf 12}, 083023 (2010).

\bibitem{jianli_prb10}
J.~Li, A.~F. Morpurgo, M.~B\"uttiker, and I.~Martin,
\newblock Marginality of bulk-edge correspondence for single-valley
  Hamiltonians,
\newblock Phys. Rev. B {\bf 82}, 245404 (2010).

\bibitem{klein99}
D.~J. Klein and L.~Bytautas,
\newblock Graphitic edges and unpaired $\pi$-electron spins,
\newblock J. Phys. Chem. A {\bf 103}, 5196 (1999).

\bibitem{liu09}
Z.~Liu, K.~Suenaga, P.~J.~F. Harris, and S.~Iijima,
\newblock Open and closed edges of graphene layers,
\newblock Phys. Rev. Lett. {\bf 102}, 015501 (2009).

\bibitem{wakabayashi09}
K.~Wakabayashi, Y.~Takane, M.~Yamamoto, and M.~Sigrist,
\newblock Edge effect on electronic transport properties of graphene
  nanoribbons and presence of perfectly conducting channel,
\newblock Carbon {\bf 47}, 124 (2009).

\bibitem{wakabayashi_jap10}
K.~Wakabayashi, S.~Okada, R.~Tomita, S.~Fujimoto, and Y.~Natsume,
\newblock Edge states and flat bands of graphene nanoribbons with edge
  modifications,
\newblock J. Phys. Soc. Jpn. {\bf 79}, 034706 (2010).

\bibitem{jaskolski11}
W.~Jaskolski, A.~Ayuela, M.~Pelc, H.~Santos, and L.~Chico,
\newblock Edge states and flat bands in graphene nanoribbons with arbitrary
  geometries,
\newblock arXiv:1104.0147v1  (2011).

\bibitem{akhmerov}
A.~R. Akhmerov and C.~W.~J. Beenakker,
\newblock Boundary conditions for Dirac fermions on a terminated honeycomb
  lattice,
\newblock Phys. Rev. B {\bf 77}, 085423 (2008).

\bibitem{zak}
J.~Zak,
\newblock Berry's phase for energy bands in solids,
\newblock Phys. Rev. Lett. {\bf 62}, 2747 (1988).

\bibitem{dahal10}
H.~Dahal, Z.~Hu, N.~Sinitsyn, K.~Yang, and A.~Balatsky,
\newblock Edge states in a honeycomb lattice: effects of anisotropic hopping
  and mixed edges,
\newblock Phys. Rev. B {\bf 81}, 155406 (2010).

\bibitem{Note1}
When $M$ is finite, there is a finite range of parameters $1 - 1/(M+1) < t'/t
  <1$, for which there are no edge states ($M$ bulk states), although the Zak
  phase is $\pi $.

\bibitem{Note2}
The energy of the edge states is zero only when the width of the system is
  larger than the localization length, otherwise the edge states at each edge
  hybridize and the resulting energy is not zero.

\bibitem{mullen}
J.~Cai, P.~Ruffieux, R.~Jaafar, M.~Bieri, T.~Braun, S.~Blankenburg, M.~Muoth,
  A.~Seitsonen, M.~Saleh, X.~Feng, K.~Mullen, and R.~Fasel,
\newblock Atomically precise bottom-up fabrication of graphene nanoribbons,
\newblock Nature Letters {\bf 466}, 470 (2010).

\bibitem{hasegawa06}
Y.~Hasegawa, R.~Konno, H.~Nakano, and M.~Kohmoto,
\newblock Zero modes of tight-binding electrons on the honeycomb lattice,
\newblock Phys. Rev. B {\bf 74}, 033413 (2006).

\bibitem{dietl08}
P.~Dietl, F.~Pi\'echon, and G.~Montambaux,
\newblock New magnetic field dependance of Landau levels in graphenelike
  structure,
\newblock Phys. Rev. Lett. {\bf 100}, 236405 (2008).

\bibitem{guinea08}
B.~Wunsch, F.~Guinea, and F.~Sols,
\newblock Dirac-point engineering and topological phase transitions in
  honeycomb optical lattices,
\newblock New J. Phys. {\bf 10}, 103027 (2008).

\bibitem{merginguniv09}
G.~Montambaux, F.~Pi\'echon, J.-N. Fuchs, and M.-O. Goerbig,
\newblock A universal Hamiltonian for motion and merging of Dirac points in a
  two-dimensional crystal,
\newblock Eur. Phys. J. B. {\bf 72}, 509 (2009).

\bibitem{delplacethese}
P.~Delplace,
\newblock {\em \'Etats de bords et c\^ones de Dirac dans des cristaux
  bidimensionnels},
\newblock PhD thesis. Universit\'e Paris-Sud XI, 2010.

\bibitem{pauling}
L.~Pauling,
\newblock Proc. N. A. S. {\bf 56}, 1646 (1966).

\bibitem{hatsugai93}
Y.~Hatsugai,
\newblock Edge states in the integer quantum Hall effect and the Riemann
  surface of the Bloch function,
\newblock Phys. Rev. B {\bf 48}, 11851 (1993).

\bibitem{hatsugaiPRL93}
Y.~Hatsugai,
\newblock Chern number and edge states in the integer quantum Hall effect,
\newblock Phys. Rev. Lett. {\bf 71}, 3697 (1993).

\bibitem{hao}
N.~Hao, P.~Zhang, Z.~Wang, W.~Zhang, and Y.~Wang,
\newblock Topological edge states and quantum Hall effect in the Haldane model,
\newblock Phys. Rev. B {\bf 78}, 075438 (2008).

\bibitem{qi06}
X.-L. Qi, Y.-S. Wu, and S.-C. Zhang,
\newblock General theorem relating the bulk topological number to edge states
  in two-dimensional insulators,
\newblock Phys. Rev. B {\bf 74}, 045125 (2006).

\bibitem{wang}
Z.~Wang, N.~Hao, and P.~Zhang,
\newblock Topological winding properties of spin edge states in Kane-Mele
  graphene model,
\newblock Phys. Rev. B {\bf 80}, 115420 (2009).

\bibitem{Note3}
Preliminary numerical results seem to confirm this prediction, (L. Bilteanu and
  C. Bena, Private communication).

\end{thebibliography}

\end{document}